\documentclass[AMA,Times1COL]{WileyNJDv5} 

\hypersetup{
    colorlinks=true,
    breaklinks,
    allcolors=blue,
    }

\usepackage{mathrsfs}
\usepackage{bm}
\usepackage{bbm}
\usepackage[capitalise]{cleveref}
\usepackage{booktabs}

\newcommand{\ie}{{\it i.e.}}

\renewcommand{\d}{\mathrm{d}}
\newcommand{\norm}[1]{\left\lVert#1\right\rVert}
\newcommand{\tens}[1]{{\boldsymbol{#1}}}

\newcommand*{\pd}[3][]{{\frac{\partial^{#1} #2}{\partial #3}}}
\newcommand{\bracs}[1]{\left({#1}\right)}
\newcommand{\bracm}[1]{\left[{#1}\right]}
\newcommand{\abs}[1]{\left|{#1}\right|}

\articletype{Original Article}%
\journal{arXiv}
\volume{1}
\copyyear{2023}
\startpage{1}

\begin{document}

\title{Gas flow and solid deformation in unconventional shale}

\author[1]{Qi Zhang}

\author[1]{Zhen-Yu Yin}

\author[2]{Xia Yan}

\author[3]{Xinyu Wang}

\authormark{ZHANG ET AL.}
\titlemark{Gas flow and solid deformation in unconventional shale}

\address[1]{\orgdiv{Department of Civil and Environmental Engineering}, \orgname{The Hong Kong Polytechnic University}, \orgaddress{\state{Hong Kong}, \country{China}}}

\address[2]{\orgdiv{School of Petroleum Engineering}, \orgname{China University of Petroleum (East China)}, \orgaddress{\state{Shandong Province}, \country{China}}}

\address[3]{\orgdiv{Department of Physics}, \orgname{University of Michigan}, \orgaddress{\state{Ann Arbor, MI 48109}, \country{United States}}}

\corres{\email{qizhang01@outlook.com}}

\abstract[Abstract]{\large Shale, a material that is currently at the heart of energy resource development, plays a critical role in the management of civil infrastructures. Whether it concerns geothermal energy, carbon sequestration, hydraulic fracturing, or waste storage, one is likely to encounter shale as it accounts for approximately 75\% of rocks in sedimentary basins. Despite the abundance of experimental data indicating the mechanical anisotropy of these formations, past research has often simplified the modeling process by assuming isotropy. In this study, the anisotropic elasticity model and the advanced anisotropic elastoplasticity model proposed by Semnani et al. (2016) and Zhao et al. (2018) were adopted in traditional gas production and strip footing problems, respectively. This was done to underscore the unique characteristics of unconventional shale. The first application example reveals the effects of bedding on apparent permeability and stress evolutions. In the second application example, we contrast the hydromechanical responses with a comparable case where gas is substituted by incompressible fluid. These novel findings enhance our comprehension of gas flow and solid deformation in shale.}


\maketitle

\makeatletter\def\Hy@Warning#1{}\makeatother
\renewcommand\thefootnote{}
\footnotetext{Preprint submitted to https://arxiv.org/}

\renewcommand\thefootnote{\fnsymbol{footnote}}
\setcounter{footnote}{1}

\section{Introduction}

\large
Shale, as an essential geomaterial with wide-ranging applications in both energy production and civil infrastructure, has garnered significant attention in recent years \cite{borja_cam-clay_2020,rezaee_fundamentals_2015,taghavinejad_flow_2020,yan_hierarchical_2021,zhang_poroelastic_2021}. Understanding the mechanical behavior of shale is crucial for optimizing extraction techniques in the energy sector and ensuring the stability of structures built on shale formations. Numerical simulation has proven to be a valuable tool in unraveling the intricate mechanics of shale, offering insights into its gas flow and solid deformation characteristics \cite{zhang_hydromechanical_2020,zhang_fluid_2021,lu_modeling_2021,yang_equivalent_2023}. This paper aims to delve into the complexities of gas flow and solid deformation in unconventional shale through comprehensive numerical investigations, shedding light on crucial factors that influence these phenomena.

Shale gas production simulation has been extensively studied especially focusing on post-hydraulic fracturing scenarios within shale gas reservoirs \cite{wang_impact_2015,yan_efficient_2018,Yan2018,yan_numerical_2019,Yan2020,yuan_numerical_2019,zhang_gas_2018}. These simulations often overlooked the directional dependence of elastic properties and the influence of bedding planes in the hydromechanical coupling \cite{shao_investigation_2020,shao_effects_2021,yin_multiscale_2023} implementation, which are inherent characteristics of shale formations \cite{borja_cam-clay_2020,ip_impacts_2021,lonardelli_preferred_2007}. Consequently, there is a notable gap in the literature regarding the investigation of field quantities predominantly influenced by these features. Addressing this research gap is of paramount importance to enhance our understanding of shale behavior and improve the accuracy of numerical simulations in shale gas production.

In addition to the study of shale gas production, this research also delves into the strip footing problem, which is a classic geotechnical inquiry \cite{zhang_strip_2021}. Traditionally, strip footing problems involve the assumption of porous material saturated with water beneath the footing. McNamee and Gibson \cite{mcnamee_displacement_1960,mcnamee_plane_1960} provided an analytical solution for such a scenario under the framework of poroelasticity. Zhang et al. \cite{zhang_preferential_2019} and Zhang et al. \cite{zhang_novel_2022} considered the double porosity nature of such porous material (shale) and investigated its preferential fluid flow patterns. Later, a multiple porosity generalization is derived in Zhang et al. \cite{zhang_mathematical_2022}. For non-linearity, Zhao and Borja \cite{zhao_continuum_2020,zhao_anisotropic_2021} extended this problem by considering plasticity effects and two effective stress measures \cite{cheng_linear_2020,cheng_intrinsic_2021}, a breakthrough in poromechanics theory. However, an intriguing question remains unexplored: What if the porous material beneath the strip footing is (nearly) saturated with a compressible gas instead? To date, no studies have investigated the implications of this scenario, and the differences between the behaviors induced by water and compressible gas remain unexplored. Thus, this study aims to address this question and discern the contrasting responses induced by these two significantly different fluids in the context of strip loading.

By examining these two prominent aspects of shale mechanics, this research seeks to advance our understanding of gas flow and solid deformation in unconventional shale. The outcomes of this study will provide valuable insights into the influence of directional dependence, bedding planes, and fluid characteristics on the behavior of shale formations. Moreover, the findings will contribute to the refinement of numerical models and improve the accuracy of simulations for shale gas production and geotechnical applications involving strip loading.

In summary, this paper will explore the intricate mechanics of gas flow and solid deformation in unconventional shale. By addressing the limitations of existing numerical simulations and investigating the influence of directional dependence, bedding planes, and fluid characteristics, our research aims to fill critical gaps in the understanding of shale behavior. The insights gained from this study will facilitate more accurate modeling and simulation approaches for shale gas production and geotechnical engineering, ultimately advancing the knowledge and practical applications in the field of unconventional shale mechanics.

\section{Verification on stress distribution}
\begin{figure}[!htb]
    \centering
    \includegraphics[width = 0.75\textwidth]{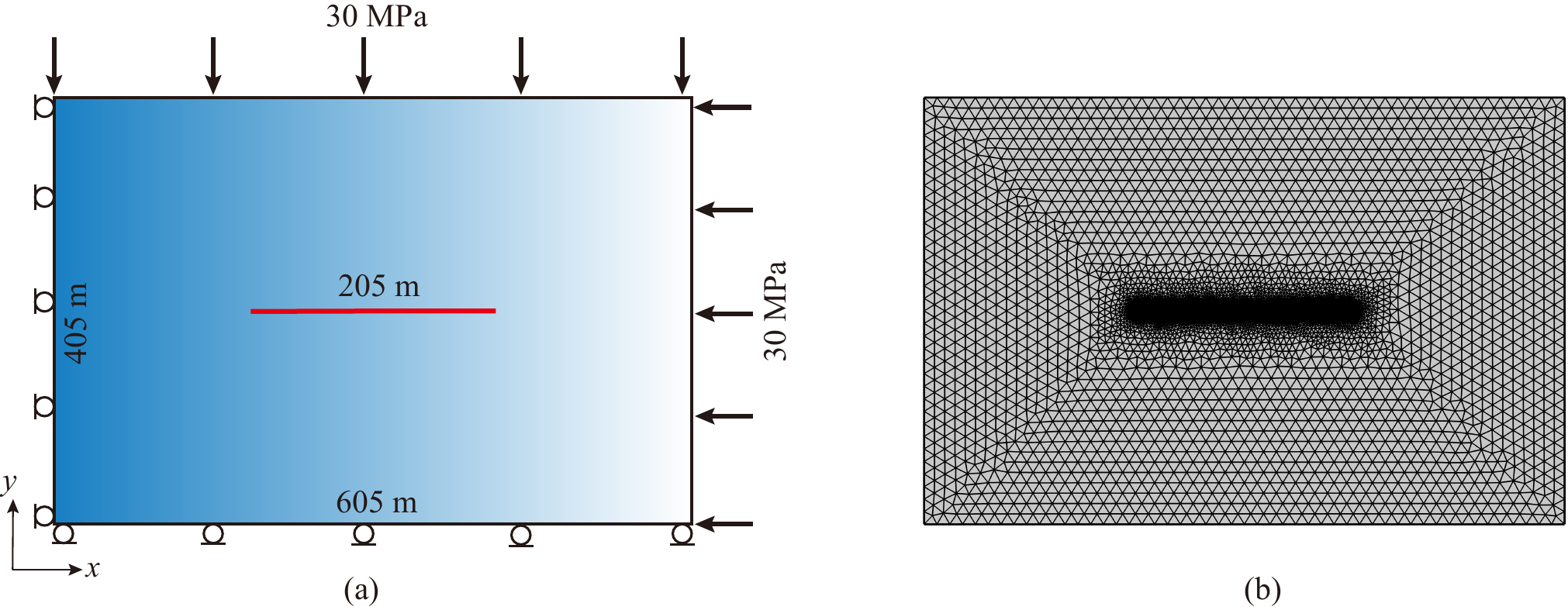}
    \caption{Schematics of (a) the reservoir model and (b) the finite element mesh with local grid refinement in COMSOL. The mechanical boundary conditions are already included in the sketch.   \label{verify_1_sk}}
\end{figure}

Before the field application, model verification is necessary. As shown by \cref{verify_1_sk}, a single fracture model is used to verify the stress distribution over the whole domain \cite{liu_efficient_2020}. The computational grid is also given in \cref{verify_1_sk}.

\begin{figure}[!htb]
    \centering
    \includegraphics[width = 0.7\textwidth]{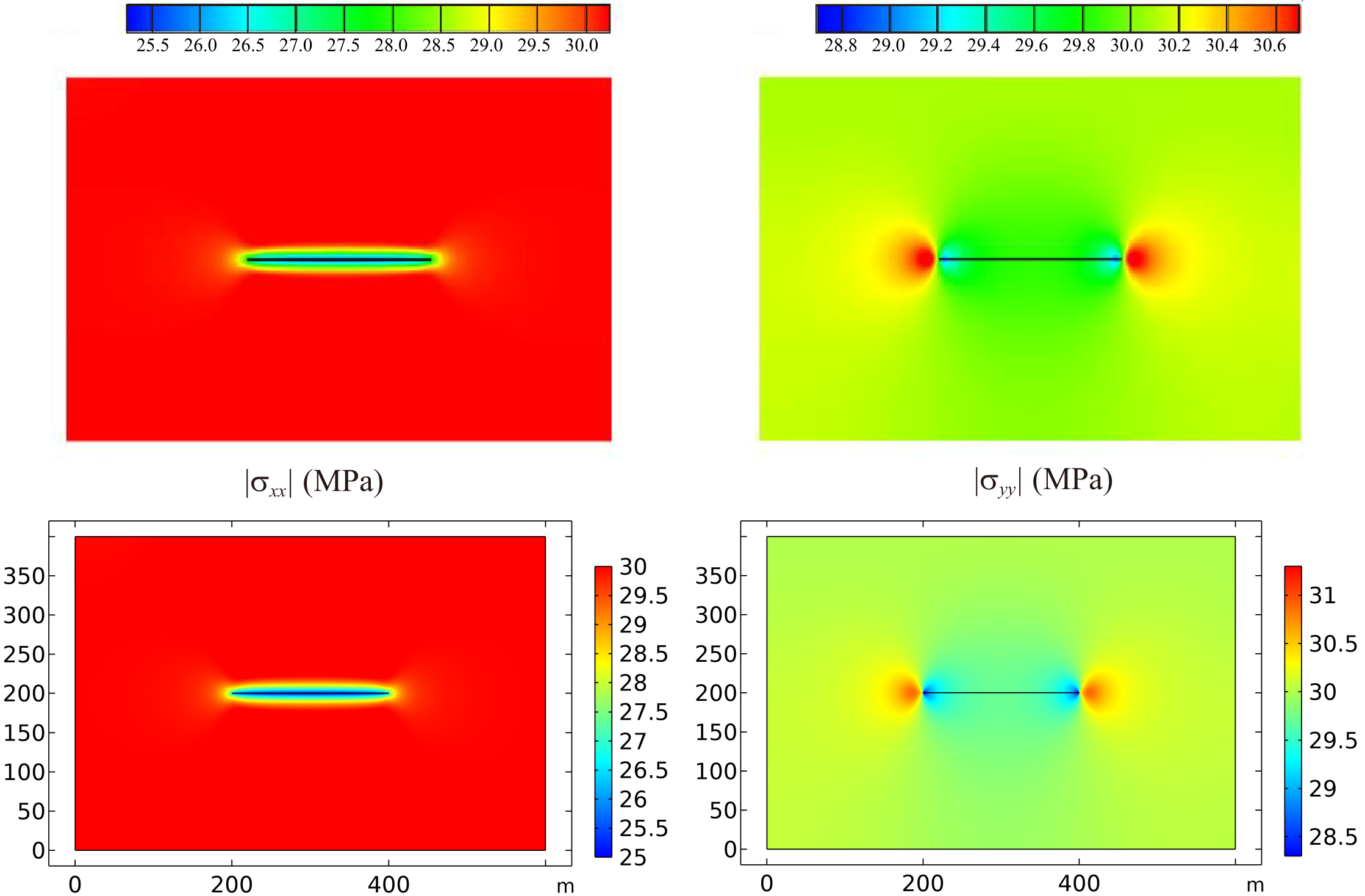}
    \caption{Comparison of $\sigma_{xx}$ and $\sigma_{yy}$ distributions calculated by the hybrid model \cite{liu_efficient_2020} (first row) and FEM in COMSOL (second row) after 100 days of production. \label{verify_1_r}}
\end{figure}

For simplicity, the fracture deformation is assumed to be elastic, and the reservoir and fracture properties are listed in \cref{verify_table1}. All the four external boundaries are impermeable, and at the center of the fracture, there is a line segment whose length is 0.2 m, and it represents a well with radius $r_w = 0.1$ m. The bottom-hole pressure (BHP) is applied on this line segment (acting as a Dirichlet boundary for fluid flow). A uniform constant compressive load of 30 MPa is applied on the right and top boundaries and maintained throughout the simulation. The left and bottom boundaries are supported by rollers. As for the initial conditions, we assume a zero displacement field and a uniform pressure field of 20 MPa, but what is more noteworthy is that for this problem, we have a non-zero effective stress field. By using the information given in \cref{verify_table1}, the initial effective stress is given as $\sigma'_{xx} = \sigma'_{yy} = -30 + 0.7\times 20 = -16$ MPa where 0.7 is the Biot coefficient. \cref{verify_1_r} shows the comparison results for $\sigma_{xx}$ and $\sigma_{yy}$ at 100 days of production, which suggests good agreement and the feasibility of our modeling method to explore stress distribution in more complex scenarios.

\begin{table}[!htb]
\centering
\caption{Parameters used in the first verification example. Note the single fracture is modeled by the material with a much lower Young's modulus. Furthermore, we use nearly incompressible fluid instead of gas in this example.}
\label{verify_table1}
\begin{tabular}{lll}
\toprule
Parameter       & Value  & Unit \\ \midrule
Matrix initial porosity   &  0.1   & 1 \\
Matrix permeability   &  0.001   & mD \\
Matrix Young's modulus  &  10  & GPa  \\ 
Matrix Poisson's ratio  & 0.2   & 1  \\
Fracture initial porosity & 0.5  &  1  \\
Fracture permeability  &  $10^4$   &  mD \\
Fracture Young's modulus & 200 & MPa \\
Fracture Poisson's ratio & 0.2 & 1 \\
Fracture initial aperture & 0.005  & m \\
Biot coefficient & 0.7 & 1 \\
Fluid compressibility & $4 \times 10^{-10}$ & 1/Pa \\
Fluid viscosity & 0.001 & $\rm Pa\cdot s$ \\
Fluid reference density & 1000 & $\rm kg/m^3$ \\
Initial pressure & 20 & MPa \\
Well radius & 0.1 & m \\
BHP & 10 & MPa \\
\bottomrule
\end{tabular} 
\end{table}

\section{Gas production analysis of anisotropic formation with discrete fracture}\label{sec3}
\subsection{Model setup}
\begin{figure}[!htb]
    \centering
    \includegraphics[width=0.65\textwidth]{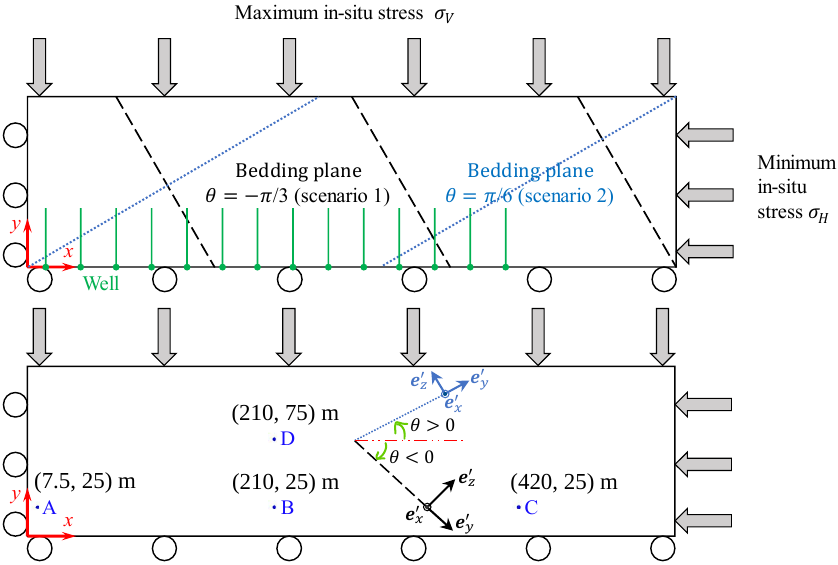}
    \caption{Schematic of a horizontal anisotropic 2D shale gas reservoir. The dashed and dotted lines represent the bedding plane orientations. No flux is applied at the outer boundaries, and BHP is prescribed at the intersection points (solid green dots) between the discrete fractures and the horizontal wellbore.}
    \label{2D_fig0}
\end{figure}
We first set up a horizontal two-dimensional (plane strain) anisotropic model as shown in \cref{2D_fig0}. For this model, we focus on the effects of bedding plane orientation and initial matrix intrinsic permeability $k_0$ on the geomechanical responses (evolutions of pressure and effective stress, gas production, etc.) of the shale gas reservoir. For the discrete fracture, the fracture flow model is adopted to describe the gas migration behavior in fracture \cite{zhao_numerical_2020,fan_analysis_2015,cao_fully_2016,liu_fully_2019,li_fully_2020}, while we ignore the impact on the solid deformation response, in other words, those green line segments in \cref{2D_fig0} are not considered in the solution of solid mechanics problem. As a result, the fracture aperture cannot be calculated from the deformation field of the fracture surface, and it is updated through the empirical relation using the fracture compressibility $c_F$ \cite{zhao_numerical_2020,cao_fully_2016}. The numerical details can be found in \cref{appx1}.

The anisotropic material parameters are based on the Trafalgar shale \cite{Cheng1997}: $E_v = 17.3$ GPa, $E_h = 20.6$ GPa, $\nu_{hh} = 0.189$, $\nu_{vh} = 0.246$, and $G_{vh} = 7.23$ GPa. We consider two scenarios of bedding plane orientation as shown in \cref{2D_fig0} by the dashed line (Scenario 1, $\theta = -\pi/3$) and the dotted line (Scenario 2, $\theta = \pi/6$). The sign convention of $\theta$ is also sketched in \cref{2D_fig0}. For Scenario 1, we further analyze three values of $k_0$, namely $5 \times 10^{-17}$ $\rm m^2$, $5 \times 10^{-18}$ $\rm m^2$, and $5 \times 10^{-19}$ $\rm m^2$. Other model parameters are given in \cref{reservoir_parameter}.

\begin{table}[!htb]
\centering
\caption{Simulation parameters for the example of gas production from an anisotropic 2D reservoir. Note the adsorption strain, adsorbed gas mass per solid volume, and surface diffusion are not considered and will be pursued in future studies. The gas apparent permeability (non-Darcy flow) follows Florence et al. \cite{florence_improved_2007}.}
\label{reservoir_parameter}
\begin{tabular}{lll}
\toprule
Parameter       & Value  & Unit \\ \midrule
Model length   &  550   & m \\
Model width   &  145   & m \\
Model thickness (in $z$ direction)  &  90   & m  \\ 
Spacing between the left boundary and the $\rm 1^{st}$ fracture & 15  &  m  \\
Fracture spacing    &   30    &  m \\  
Fracture length   &   50   & m \\ 
Initial fracture aperture $w_0$    &  $5 \times 10^{-3}$   & m \\
Initial fracture porosity $\phi_{F0}$    &   0.3   &  1 \\
Initial fracture permeability $k_{F0}$   & $10^{-12}$   &  $\rm m^2$  \\
Fracture compressibility $c_F$   &   0.01     &  $\rm MPa^{-1}$   \\  
Initial reservoir pressure     &   20   &  MPa \\
Bottom-hole pressure (BHP)   &  3.5    &  MPa \\
Maximum in-situ horizontal stress $\sigma_V$   &   -40    & MPa  \\
Minimum in-situ horizontal stress $\sigma_H$   &   -35    & MPa  \\
Initial matrix intrinsic permeability $k_0$  &  $5\times 10^{-19}$   &  $\rm m^2$ \\
Initial matrix porosity $\phi_0$   &   0.05  &  1 \\
Bulk modulus of the solid grain $K_s$    & 45    & GPa \\
Gas viscosity $\mu_g$    &  $2\times 10^{-5}$   &   $\rm Pa\cdot s$ \\
Reservoir temperature $T$   &  353.15  & K \\
Gas molar mass $M_g$ & 16.04 & $\rm g/mol$ \\
Langmuir pressure $P_L$ & 4  & MPa\\ \bottomrule
\end{tabular} 
\end{table}

\subsection{Role of initial matrix intrinsic permeability}

\begin{figure}[!htb]
\centering
	\includegraphics[width = 0.8\textwidth]{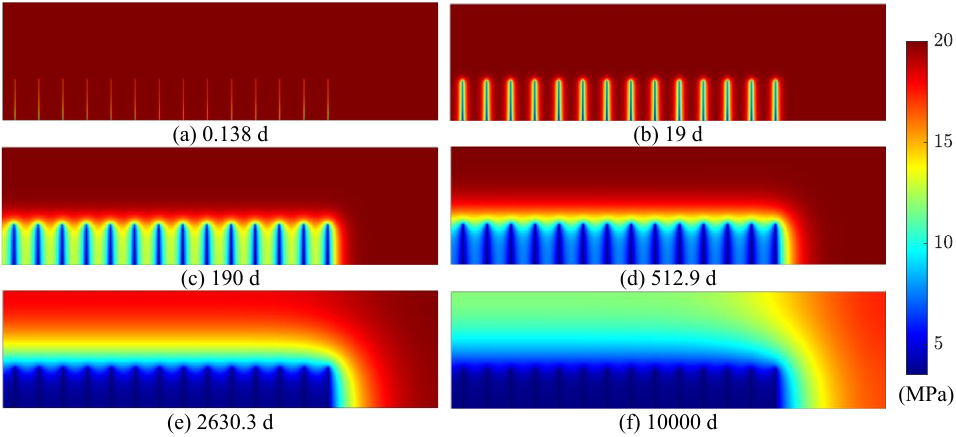}
	\caption
	{\label{Shalegas_fig2}
Evolution of gas pressure distribution in Scenario 1 when $k_0 = 5 \times 10^{-19}$ $\rm m^2$. For this figure, we focus on the propagation pattern of $p_g$ and compare with \cref{Shalegas_fig3}.}
\end{figure}

\begin{figure}[!htb]
\centering
	\includegraphics[width = 0.8\textwidth]{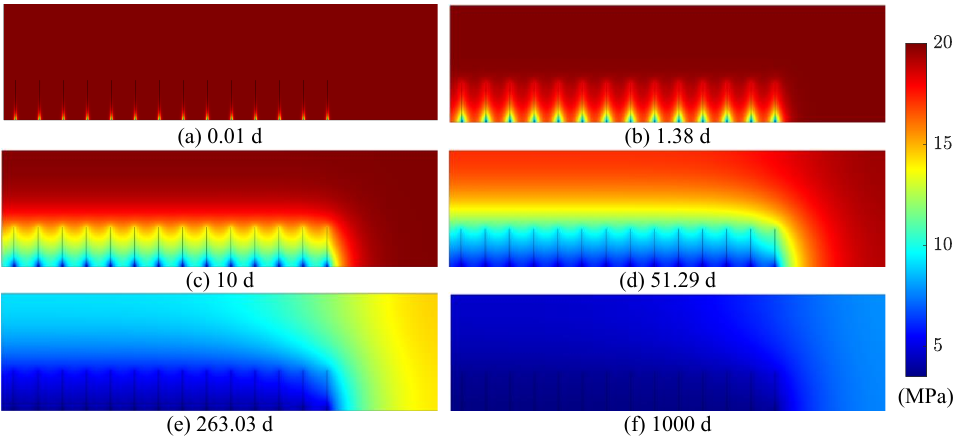}
	\caption
	{\label{Shalegas_fig3}
Evolution of gas pressure distribution in Scenario 1 when $k_0 = 5 \times 10^{-17}$ $\rm m^2$. One could clearly see the propagation pattern difference by comparing with \cref{Shalegas_fig2}. Note the similar phenomenon would be observed if we fix $k_0$ while decreasing $k_{F0}$ \ie, decrease the permeability contrast.}
\end{figure}

\cref{Shalegas_fig2} and \cref{Shalegas_fig3} present the evolution of gas pressure $p$ under different $k_0$. It can be observed that the minimum pressure always occurs around the discrete fracture. However, the diffusion pattern changes when we increase $k_0$. From \cref{Shalegas_fig2}, we can see that when $k_0 = 5 \times 10^{-19}$ $\rm m^2$, the pressure propagates from the whole segment of the discrete fracture. In other words, the discrete fracture acts as a line sink. In contrast, when $k_0$ is increased to $5 \times 10^{-17}$ $\rm m^2$ in \cref{Shalegas_fig3}, we can observe a pressure radiation pattern surrounding the intersection point between the discrete fracture and the horizontal well, similar to a point sink.
% \begin{figure}[!htb]
%     \centering
%     \includegraphics[scale=0.6]{Fig_apply1/gas_mass_k0.png}
%     \caption{Impact of initial matrix intrinsic permeability $k_0$ on the gas production behaviors. }
%     \label{gasmass}
% \end{figure}

\subsection{Impacts of anisotropy on pressure and permeability}
\begin{figure}[!htb]
    \centering
    \includegraphics[width = 0.8\textwidth]{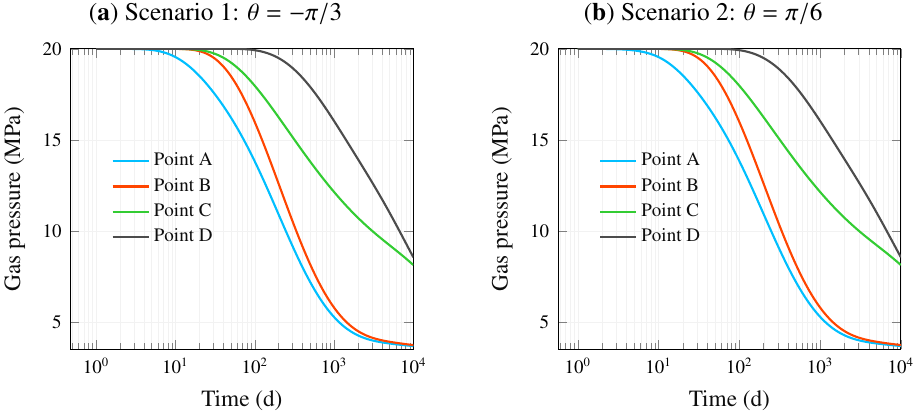}
    \caption{Gas pressure decay at four investigation points, which suggests that the elastic anisotropy has negligible influences, as the ``driven force'' comes from the gas compressibility itself.}
    \label{fig:newadd_p}
\end{figure}

\begin{figure}[!htb]
    \centering
    \includegraphics[width = 0.8\textwidth]{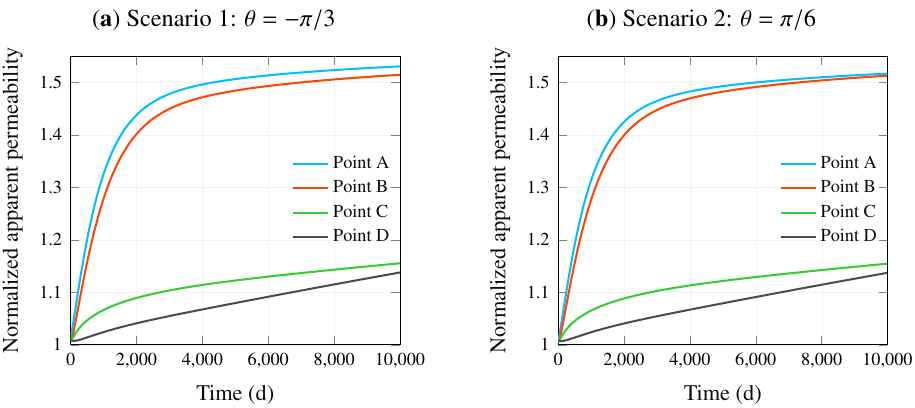}
    \caption{Apparent permeability evolution at four investigation points. Due to the ``boundary effect'' on Points A and B,  when we change $\theta$, the porosity evolution is affected, which in turn affects the apparent permeability.}
    \label{fig:newadd_ka}
\end{figure}

Now we discuss the effect of bedding plane orientation. In other words, we want to explore the directional dependence of the hydromechanical responses. While suggested by \cref{fig:newadd_p}, we find the gas pressure $p$ distribution is not sensitive to the bedding plane orientation. Instead, the anisotropy affects the permeability evolution. In \cref{fig:newadd_ka}, we show the evolution of apparent permeability at the same four investigation points (see \cref{2D_fig0} for their locations). Now the difference is obvious, the green and red curves get closer in Scenario 2 than in Scenario 1. This reminds us of the fact that the apparent permeability is not only a function of gas pressure but also depends on the porosity change, and the anisotropy could directly control $\phi$. As mentioned in Chin et al. \cite{chin_fully_2000}, the change of effective stress would affect the permeability of the matrix, so we would expect to see some differences in the pattern of $\tens{\sigma}'$ with different $\theta$, as discussed next.

\subsection{Effective stress evolution}
\begin{figure}[!htb]
\centering
	\includegraphics[width = 0.8\textwidth]{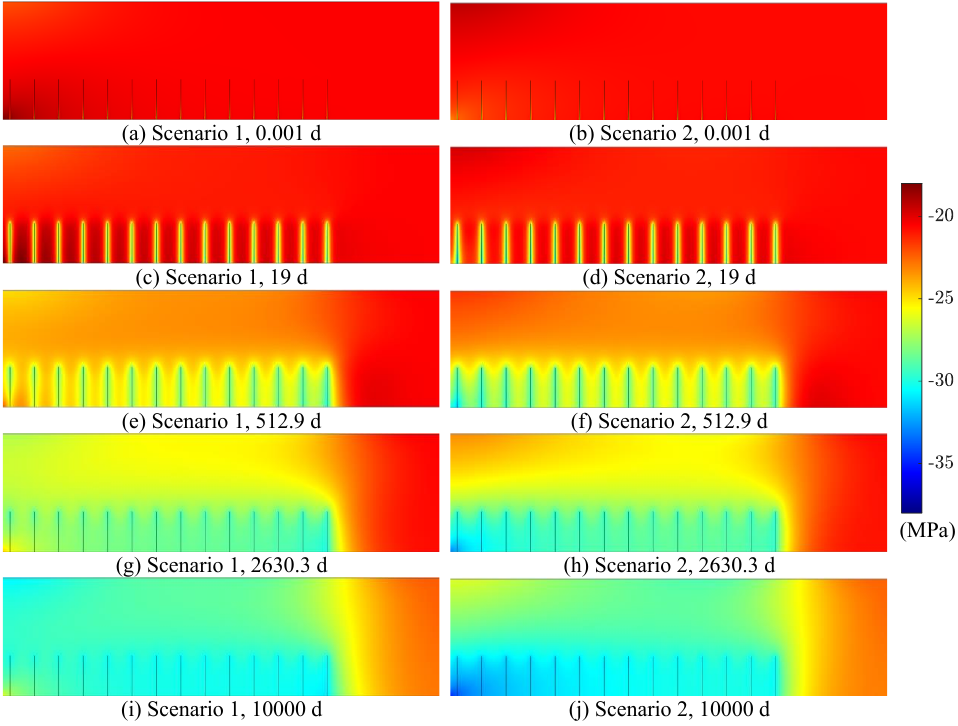}
	\caption
	{\label{Shalegas_fig5}
Comparison of the effective stress component $\sigma'_{xx}$ distribution evolution between Scenarios 1 and 2. The differences are attributed to both elastic anisotropy and ``boundary constraint effect''.}
\end{figure}

\begin{figure}[!htb]
\centering
	\includegraphics[width = 0.8\textwidth]{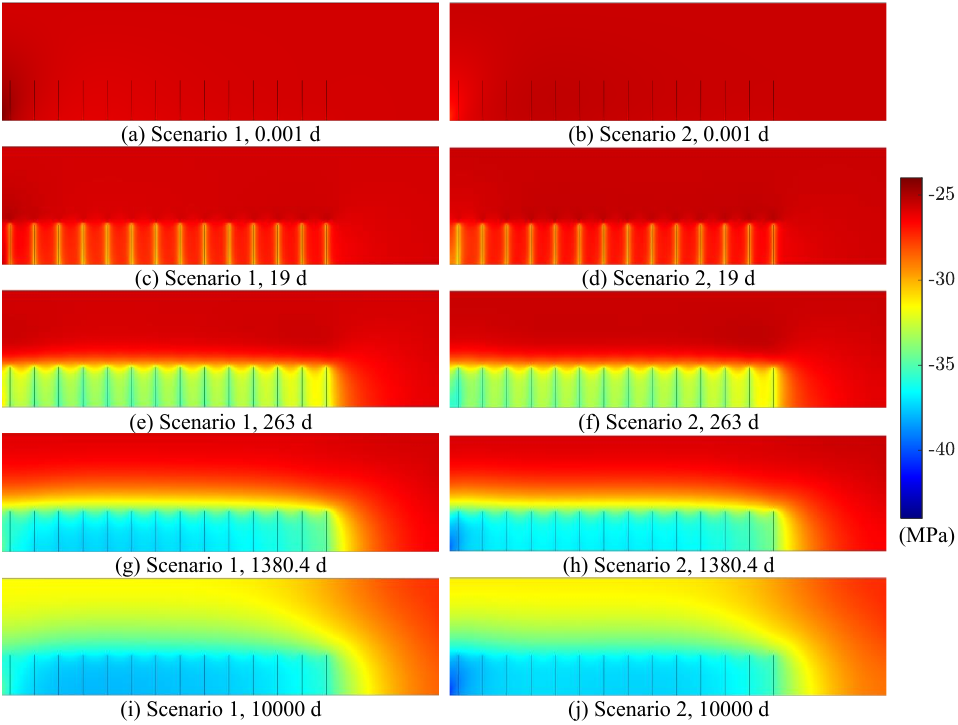}
	\caption
	{\label{Shalegas_fig6}
Comparison of the effective stress component $\sigma'_{yy}$ distribution evolution between Scenarios 1 and 2. The differences are mainly attributed to the ``boundary constraint effect''.}
\end{figure}

Next, the impact of anisotropy on the effective stress $\bm{\sigma}'$ distribution is explored. \cref{Shalegas_fig5} and \cref{Shalegas_fig6} show the comparisons of $\sigma'_{xx}$ and $\sigma'_{yy}$ between Scenario 1 and Scenario 2, respectively. Consequently, transversely isotropic characteristics can have a considerable impact on reservoir stress changes \cite{zhu_4d_2018}. For $\sigma'_{xx}$, the difference is localized around the lower-left and upper-left corners. For example, the absolute value $\abs{\sigma'_{xx}}$ at the lower-left corner is smaller than the surrounding areas in Scenario 1 when $\theta = -\pi/3$; while it is the opposite in Scenario 2 when%
\begin{figure}[!htb]
    \centering
    \includegraphics[scale=0.6]{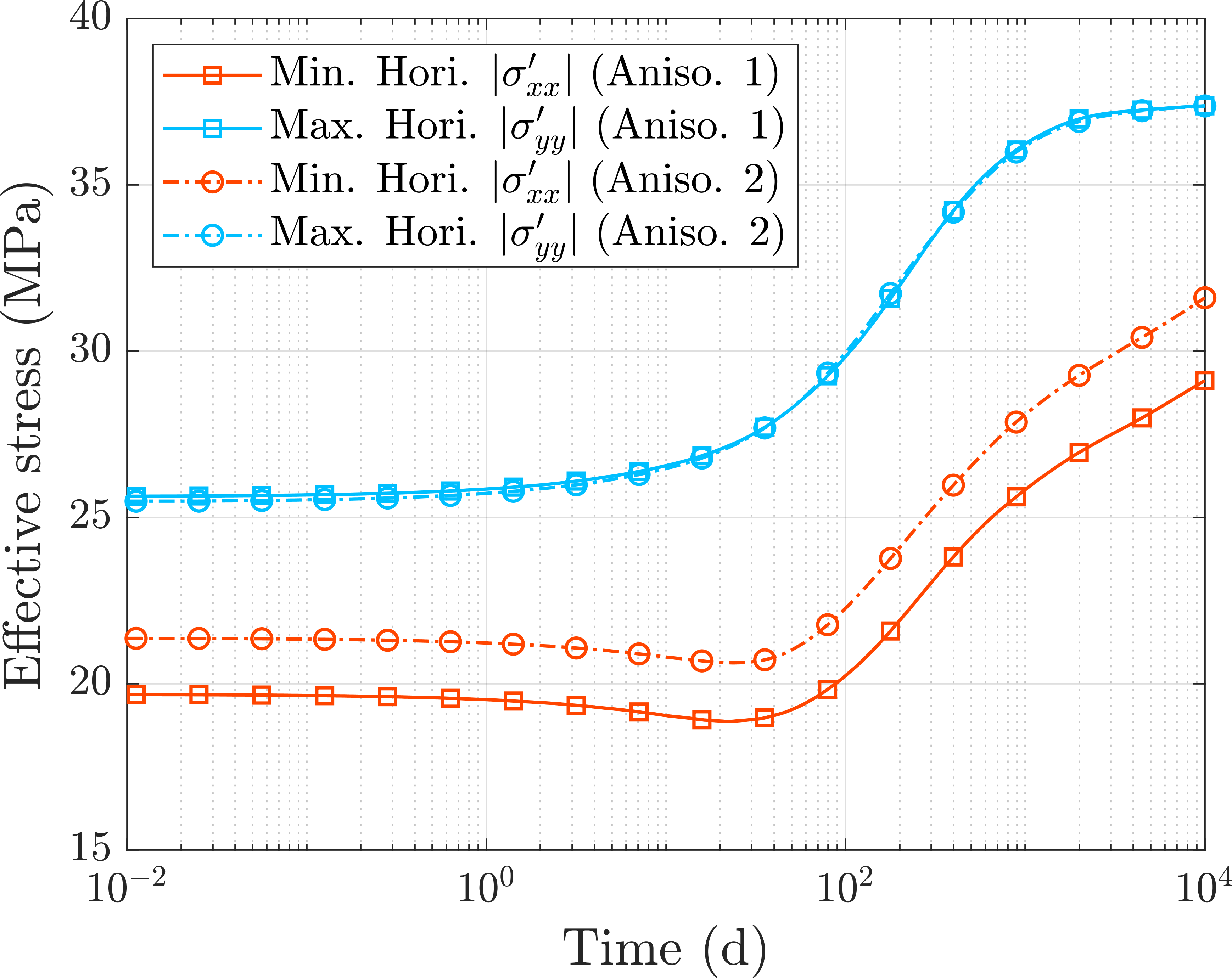}
    \caption{Effect of different anisotropy scenarios (Aniso. 1: $\theta = -\pi/3$ and Aniso. 2: $\theta = \pi/6$) on minimum and maximum horizontal effective stress evolutions at (60, 25) m. \label{newfig1}}
\end{figure}
\noindent $\theta = \pi/6$. This could explain the change of the blue curves between Scenarios 1 and 2 in \cref{fig:newadd_ka}. In Scenario 1, this corner region with low  $\abs{\sigma'_{xx}}$ renders the stress propagation pattern of the first discrete fracture to be different from the remaining discrete fractures (\cref{Shalegas_fig5}e), and it also leads to several dark red regions between adjacent discrete fractures (\cref{Shalegas_fig5}c), demonstrating a certain level of heterogeneity. In Scenario 2, the aforementioned features of Scenario 1 are less obvious, but in the later period, the contour line (especially the yellow line) is substantially affected by this change in $\theta$ (\cref{Shalegas_fig5}i and \cref{Shalegas_fig5}j). For $\sigma'_{yy}$, the difference between Scenario 1 and Scenario 2 is only significant in the later period at the lower-left corner (\cref{Shalegas_fig6}g to \cref{Shalegas_fig6}j). In other words, the anisotropy has a stronger effect on $\sigma'_{xx}$ than $\sigma'_{yy}$. Furthermore, if we compare the propagation patterns of $p$, $\sigma'_{xx}$, and $\sigma'_{yy}$, we may notice that the patterns of $p$ and $\sigma'_{yy}$ are quite similar to each other since they both spread from the stimulated reservoir domain (SRD) to the non-stimulated reservoir domain (NSRD), instead, for $\sigma'_{xx}$, it changes simultaneously in SRD and NSRD. While \cref{Shalegas_fig5} and \cref{Shalegas_fig6} show the overall trend, in \cref{newfig1}, we plot the effective stress at one observation point (60, 25) m, in which additional useful information can be captured. The bedding plane orientation has an evident effect on the minimum horizontal effective stress evolution, but its effect on the maximum horizontal effective stress evolution is negligible. In addition, at $t = 10^4$ d, the blue curve is almost flat, while the red curves are still increasing, which indicates the occurrence of stress re-orientation during depletion \cite{tang_geomechanics_2022}. Furthermore, the evolution of the minimum horizontal effective stress is not monotonic at this selected point and its surrounding region.

\subsection{Differences between isotropy and transverse isotropy}
\begin{figure}[!htb]
    \centering
    \includegraphics[width=0.8\textwidth]{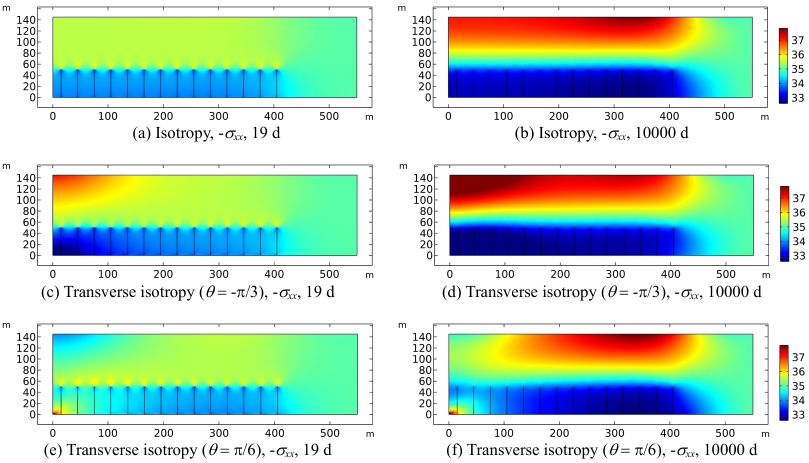}
    \caption{Comparison of total stress $\sigma_{xx}$ (MPa) for isotropic and transversely isotropic elastic models. The differences are attributed to both elastic anisotropy and ``boundary constraint effect''. \label{newfig2}}
\end{figure}

\begin{figure}[!htb]
    \centering
    \includegraphics[width=0.8\textwidth]{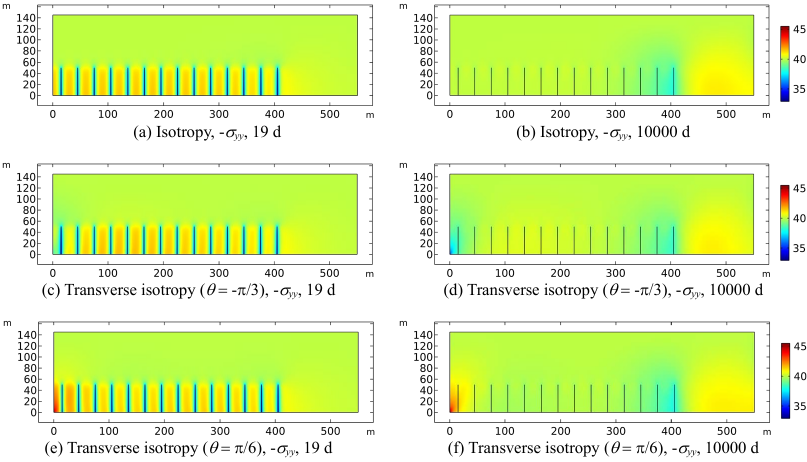}
    \caption{Comparison of total stress $\sigma_{yy}$ (MPa) for isotropic and transversely isotropic elastic models. The differences are mainly attributed to the ``boundary constraint effect''. Thus the choice of this example may not be ideal. In principle, we should have a core region that is far away from all the boundaries so that the boundary conditions do not affect the field quantities of interest. \label{newfig3}}
\end{figure}

Finally, we compare the total stress $\tens{\sigma}$ between isotropic and transversely isotropic models, similar to the process in Tang et al. \cite{tang_geomechanics_2022}. In this study, we assign $E = 18$ GPa and $\nu = 0.25$ to the isotropic model, as these parameters ensure the same generalized bulk modulus with the anisotropic model \cite{Cheng1997}. \cref{newfig2} and \cref{newfig3} portray the results at two different time slots ($t = 19$ d and $t = 10000$ d) for isotropic and transversely isotropic elastic models. First of all, it is easy to check that the result is consistent with the traction boundary condition, \ie, $\abs{\sigma_{xx}} = 35$ MPa on the right boundary and $\abs{\sigma_{yy}} = 40$ MPa on the top boundary. Second, for isotropic model, our result is consistent with the pattern in Liu et al. \cite{liu_efficient_2020}. For example, we can observe the stress ($\sigma_{xx}$) concentration at the fracture tip that exhibits a bulb shape, similar to that in \cref{verify_1_r}, $\abs{\sigma_{xx}}$ decreases in the depleted area and increases on the top of the domain, and there are many narrow bands with high values of $\abs{\sigma_{yy}}$ between adjacent fractures in the early production stage. Third, for anisotropic model, the influence range of $\sigma_{xx}$ is much larger than that of $\sigma_{yy}$. As shown in \cref{newfig2}, the anisotropy could affect the $\sigma_{xx}$ distribution up to 150 m, while in \cref{newfig3}, the difference is only localized around the lower-left corner. This finding is quite consistent with the findings from the work on rock anisotropy when estimating in-situ stresses \cite{crawford_determining_2020,amadei_importance_1996,khan_impact_2012}, \ie, as opposed to the habitually applied isotropic assumption, the minimum horizontal stress $S_{h,\,\min}$ shows a strong dependency on the anisotropic poroelastic properties, and this $S_{h,\,\min}$ would control simulated hydraulic fracture geometries and proppant concentration \cite{khan_impact_2012,crawford_determining_2020}. Nevertheless, for both $\sigma_{xx}$ and $\sigma_{yy}$, the formation anisotropy increases the stress distribution heterogeneity, which is unfavorable for wellbore stability and sustainable production \cite{asaka_anisotropic_2021}.

\section{Strip load on a gas-saturated elastoplastic porous medium}

\begin{figure}[!htb]
	\begin{center}
	\begin{tabular}{c}
	\includegraphics[width = 0.7\textwidth]{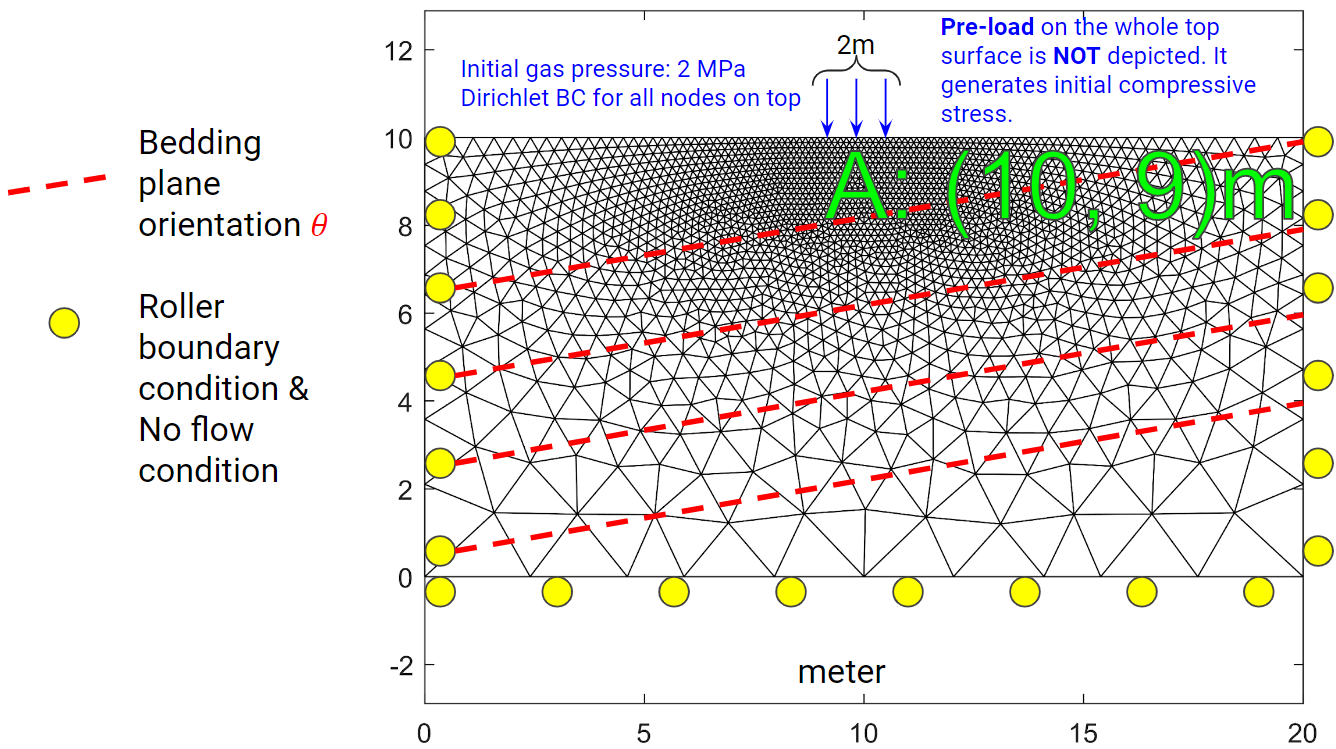}
	\end{tabular}
	\end{center}
	\caption
	{\label{sfmesh}
Schematic of the gas-saturated transversely isotropic porous medium subjected to a strip load. Gravity is ignored in this example. Note that there is a uniform preloading of 22 MPa.}
\end{figure}

We now introduce anisotropy in both the elastic and plastic behaviors through the advanced model proposed by Semnani et al. \cite{semnani_thermoplasticity_2016} and Zhao et al. \cite{zhao_strength_2018}. We conduct a plane strain simulation over a rectangular domain of 20 m $\times$ 10 m subjected to a central strip load. The domain is assumed to be transversely isotropic, a type of anisotropy exhibited by many natural materials, and it is always represented by inclined bedding planes where the angle between the bedding plane and the horizontal direction is $\theta$, as shown in \cref{sfmesh}. Through this example, we demonstrate how the solutions change with the following factors: (a) non-Darcy flow \cite{florence_improved_2007}; (b) bedding plane orientation $\theta$; (c) stress history. The uniform pre-load is denoted as $\omega_0 = 22$ MPa, and gravity is ignored in this example. To make the non-Darcy flow more prominent, we assume an initial gas pressure equal to 2 MPa. In the beginning, a strip load $\omega = 20$ MPa is applied in a very short period over a width of 2 m, which makes the domain globally undrained. The strip load is then held constant for the remaining of the simulation. Mechanical and flow parameters used in the simulation are given as follows: $E_v = 12858$ MPa, $E_h = 21900$ MPa, $G_{vh} = 6510$ MPa, $\theta = \pi/4$, $\lambda^p = 0.0013$, $c_1^p = 0.7$, $c_2^p = -0.36$, $c_3^p = 0.6$, $M_{\rm critical} = 1.07$, $M_g = 16.04$ g/mol, $p_{c0} = -40$ MPa, $K_s = +\infty$, $\phi_0 = 0.04$, $k_0 = 2\times 10^{-20}$ $\rm m^2$, $\mu_g = 2\times 10^{-5}$ $\rm Pa \cdot s$, $T = 293.15$ K, $\epsilon_L = 0$, and $P_L = +\infty$.

The simulation goes as follows. Under the pre-load $\omega_0$, the initial (effective) stress field in $x$, $y$, and $z$ directions are -4 MPa, -20 MPa, and -4 MPa, respectively. The strip load $\omega$ is then applied in 10 loading steps with $\Delta t$ = 1 second, thus generating a transient solution of excess gas pressure. The transient solution continues for additional 50 time steps while holding $\omega$ constant. To simulate the whole process, the time increment is increased by a factor of 1.25 from the previous value, \ie, $\Delta t_{n+1} = 1.25 \Delta t_n$ ($n = 1, 2, \ldots, 49$) and $\Delta t_1 = 31.25$ min. Here, the excess gas pressure is the gas pressure increment from its initial value of 2 MPa. The source code can be downloaded from this \href{https://gitee.com/qzhang94/source-code-arXiv-2311.12877}{code repository}.

\begin{figure}[!htb]
	\begin{center}
    \includegraphics[width=0.75\textwidth]{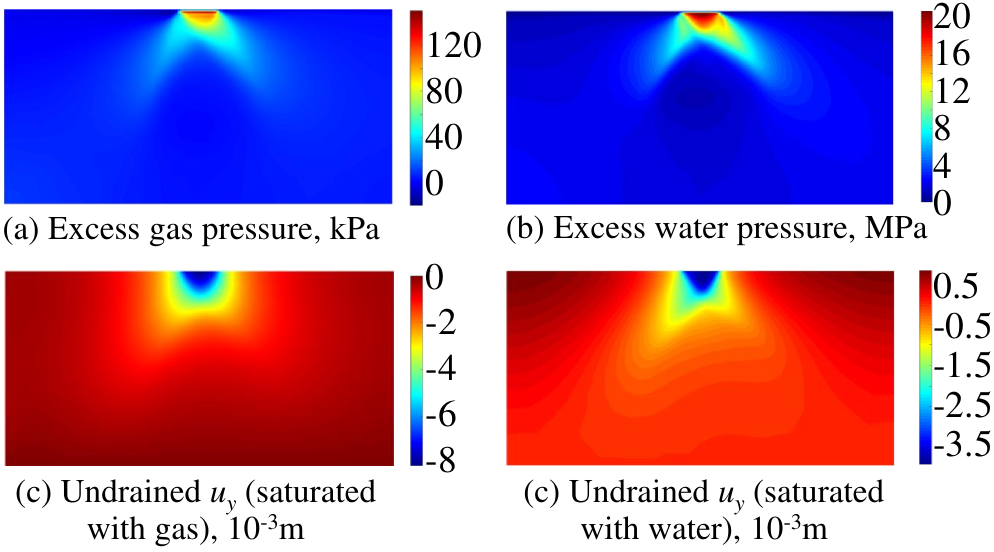}
	\end{center}
	\caption
	{\label{newcase_01} Contours of excess pressure and undrained vertical displacement $u_y$ at $t = 10$ s (right after the strip load $\omega$ is applied). Constant permeability $k \equiv k_0$ is assumed for both water and gas flows. Left: Gas-saturated porous medium. Right: Water-saturated porous medium.}
\end{figure}

We first make some quantitative comparisons with the results for slightly compressible fluids such as water \cite{zhao_continuum_2020,zhang_mathematical_2022} of the same strip load problem. For liquid, we can set the initial pressure to 0 and pre-load is therefore $\omega_0 = 20$ MPa, in order to maintain the same initial (effective) stress field. An obvious difference is that for gas or highly compressible fluid, the induced pressure increment is much less than the applied load magnitude $\omega$, as shown in \cref{newcase_01}a. In poroelasticity theory, this is known as the Skempton effect \cite{Wang2000}, and for highly compressible fluid constituent, the Skempton coefficient $B$ is approaching zero \cite{Wang2000}. We can see that this is also true for poroelastoplasticity. In other words, the porous material behaves as an elastoplastic material without fluid, and that would explain why the undrained deformation in \cref{newcase_01}c is larger than that in \cref{newcase_01}d. These qualitative consistencies between the numerical simulation and the mathematical theory indicate the correctness of our code implementation. However, we should remark that the plastic deformation of water-saturated porous medium could be greater than the gas-saturated porous medium.

\begin{figure}[!htb]
	\begin{center}
	\begin{tabular}{c}
	\includegraphics[width = 0.5\textwidth]{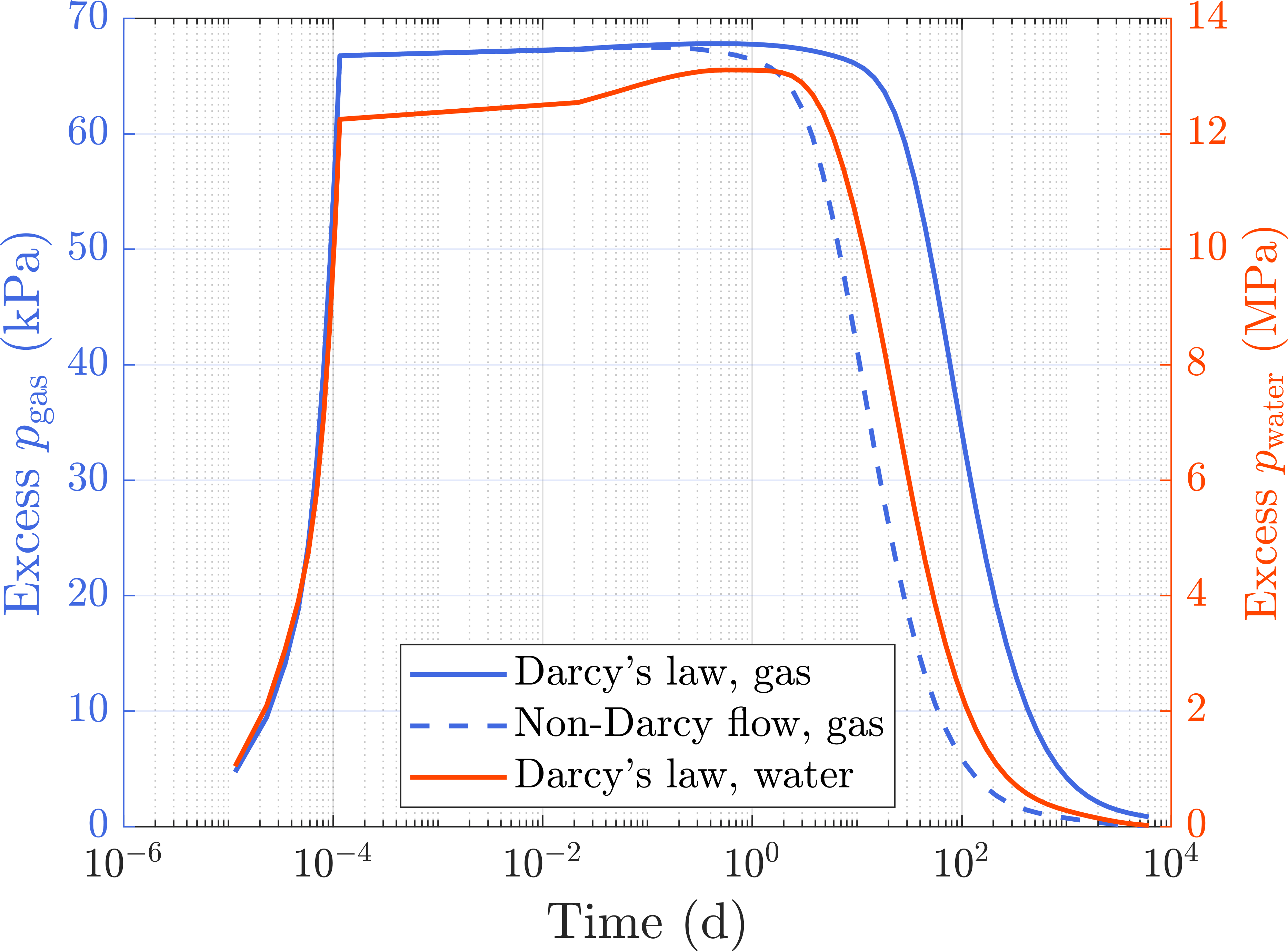}
	\end{tabular}
	\end{center}
	\caption
	{\label{newcase_02}
Evolution of excess pressure at point A with different permeability models and different kinds of liquids. The Mandel-Cryer effect is not significant on gas, and the Non-Darcy flow \cite{florence_improved_2007} effect is significant in the dissipation stage.}
\end{figure}

Due to the large undrained deformation, in the remaining 50 time steps, the ``driven force'' of excess gas pressure dissipation mainly comes from the compressibility of the gas itself, rather than the compression of pore spaces. This could explain why the so-called Mandel-Cryer effect is not prominent in this example, as shown in \cref{newcase_02}. In addition, since the compressibility of gas is much higher than that of water, even though the gas mobility is higher than that of water, the dissipation time is still comparable to that of water, as shown in \cref{newcase_02}. In \cref{newcase_02}, we also compare the dissipation of excess gas pressure between Darcy and non-Darcy flows, which suggests that non-Darcy flow would enhance the dissipation process because $k_{a}$ is nearly 10 times of $k_0$ in this example. We can imagine that if we also include surface diffusion, the descending segment of the blue dashed curve will continue moving to the left.

\begin{figure}[!htb]
	\begin{center}
    \includegraphics[width=0.75\textwidth]{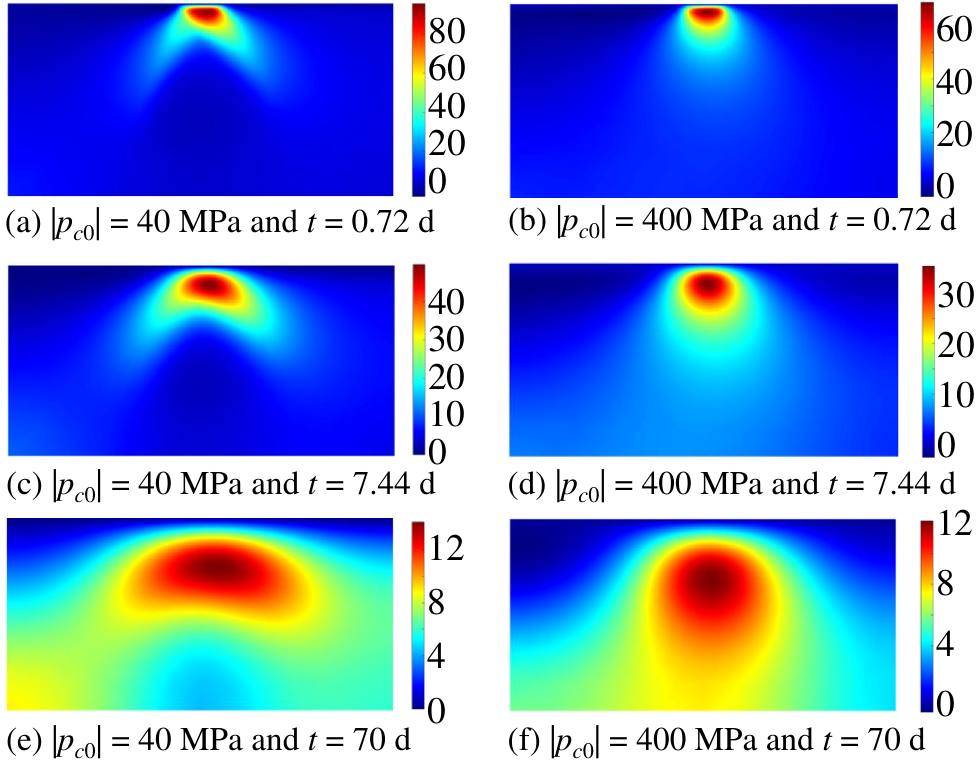}
	\end{center}
	\caption
	{\label{newcase_06} Contours of excess gas pressure at three different time slots. Left: $p_{c0} = -40$ MPa (elastoplastic deformation). Right: $p_{c0} = -400$ MPa (elastic deformation). Color bars are pressures in kPa.}
\end{figure}

\begin{figure}[!htb]
	\begin{center}
	\begin{tabular}{ccc}
	\includegraphics[width = 0.3\textwidth]{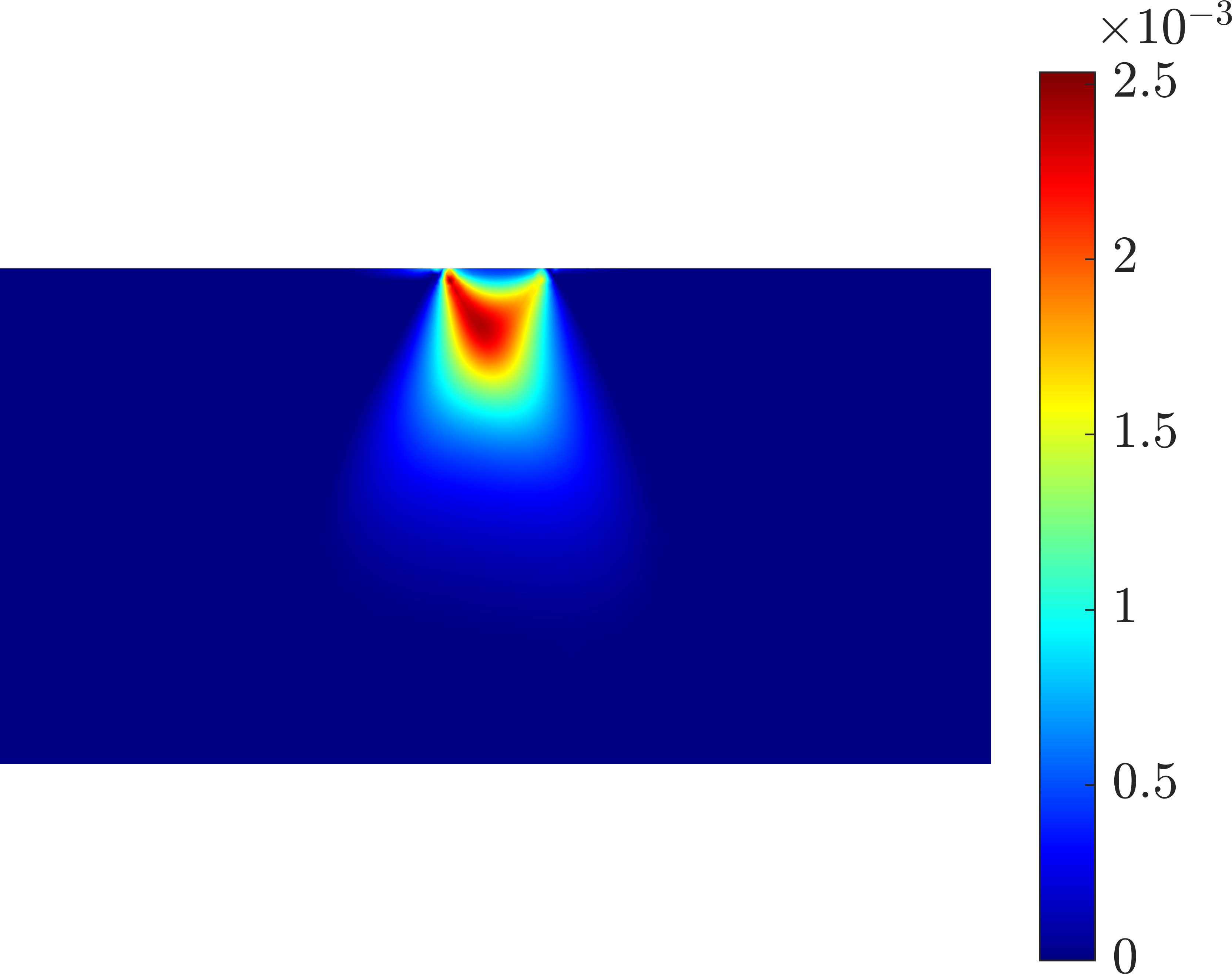} & \includegraphics[width = 0.3\textwidth]{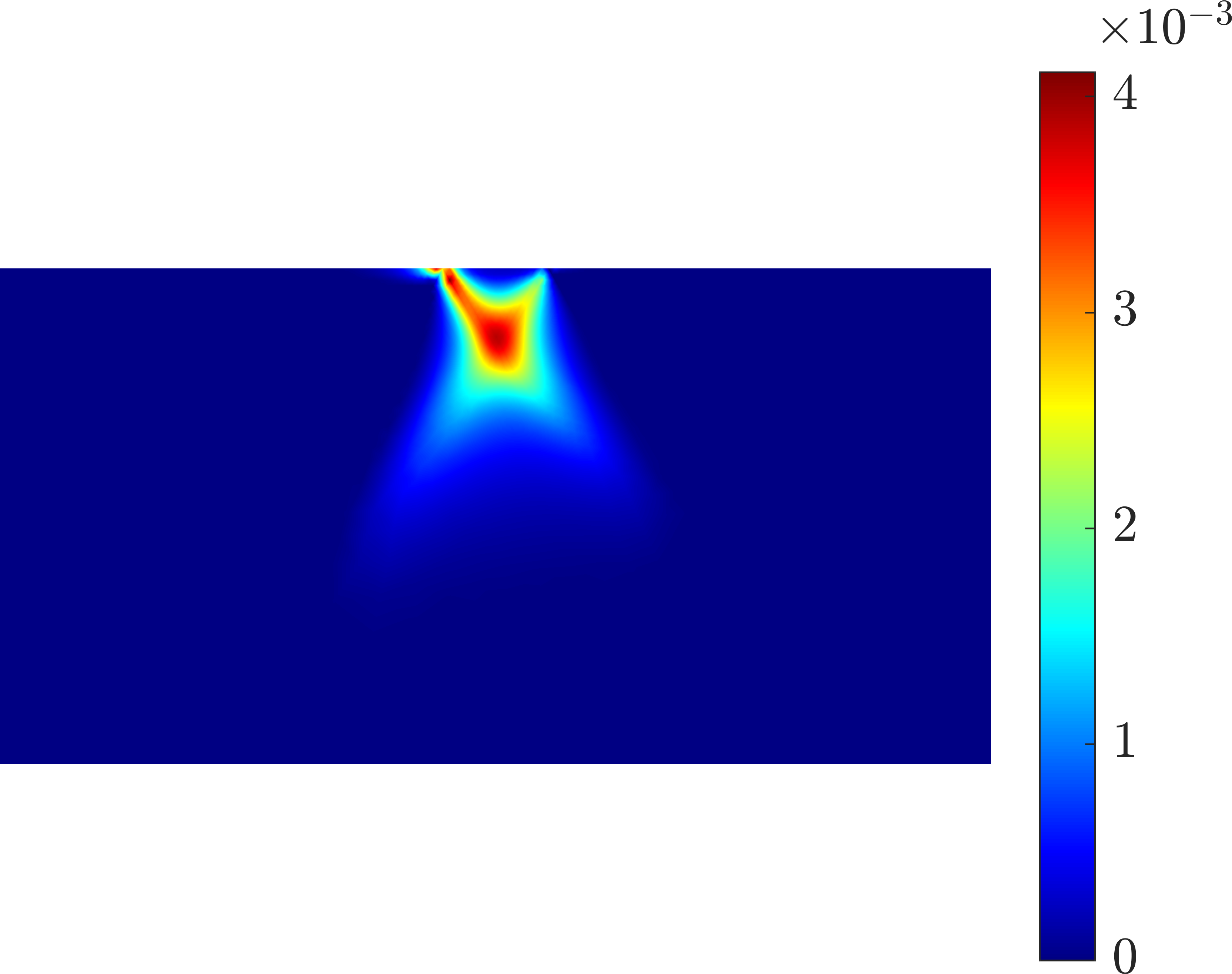} & \includegraphics[width = 0.3\textwidth]{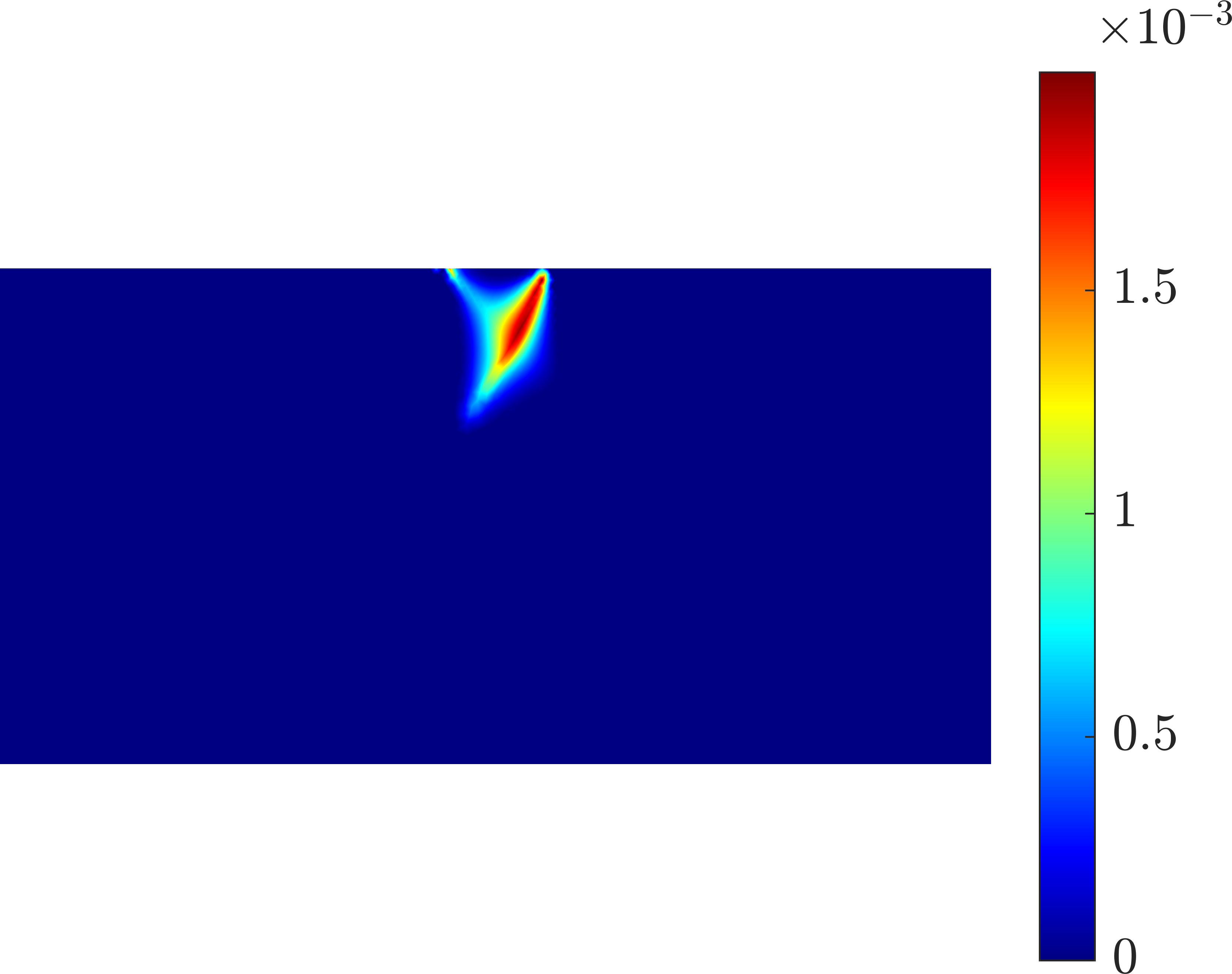}\\
	(a) $\norm{\tens{\epsilon}^p}$ when $\theta = \pi/15$  & (b) $\norm{\tens{\epsilon}^p}$ when $\theta = \pi/4$ & (c) $\norm{\tens{\epsilon}^p}$ when $\theta = 7\pi/18$  \\
	\includegraphics[width = 0.3\textwidth]{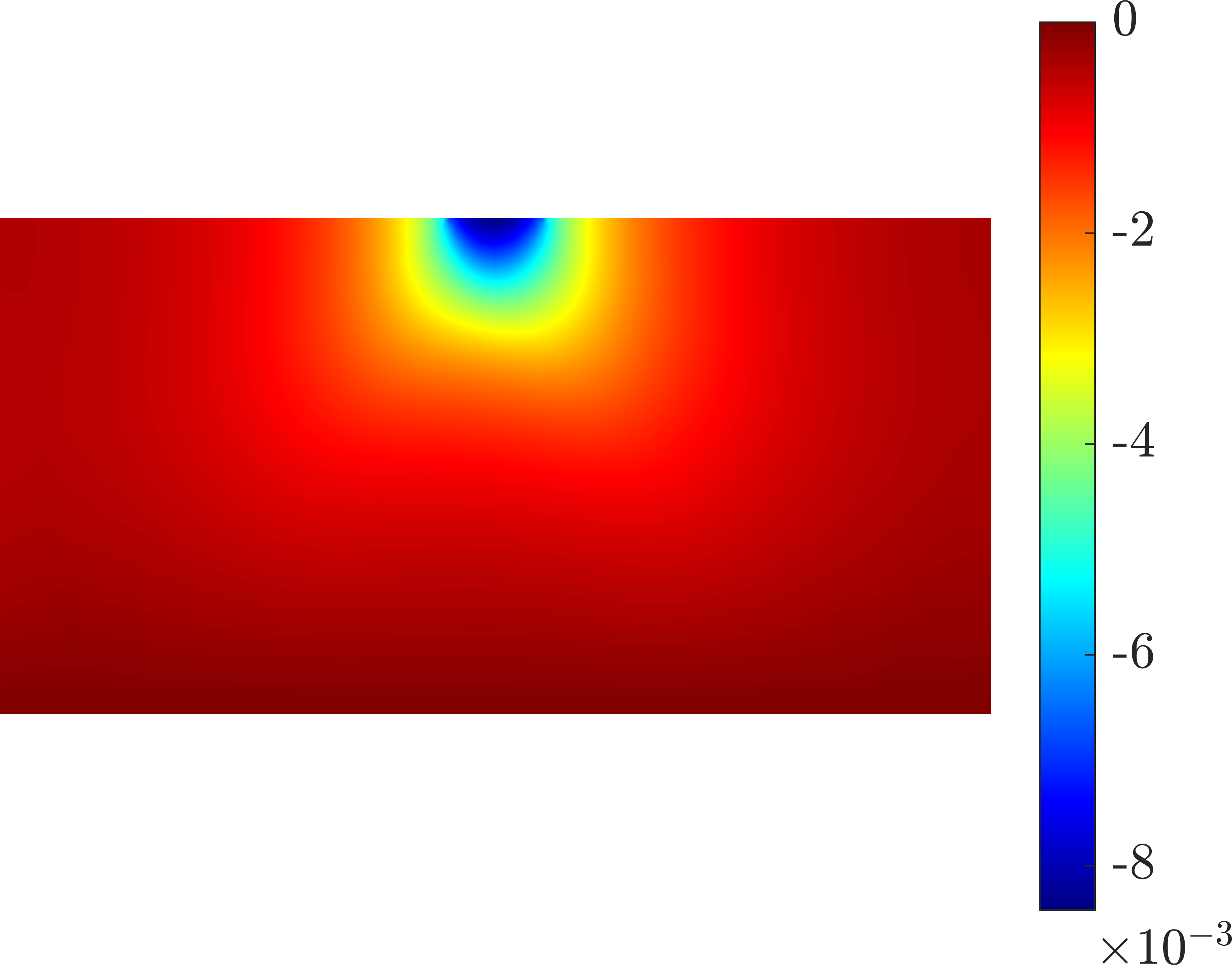} & \includegraphics[width = 0.3\textwidth]{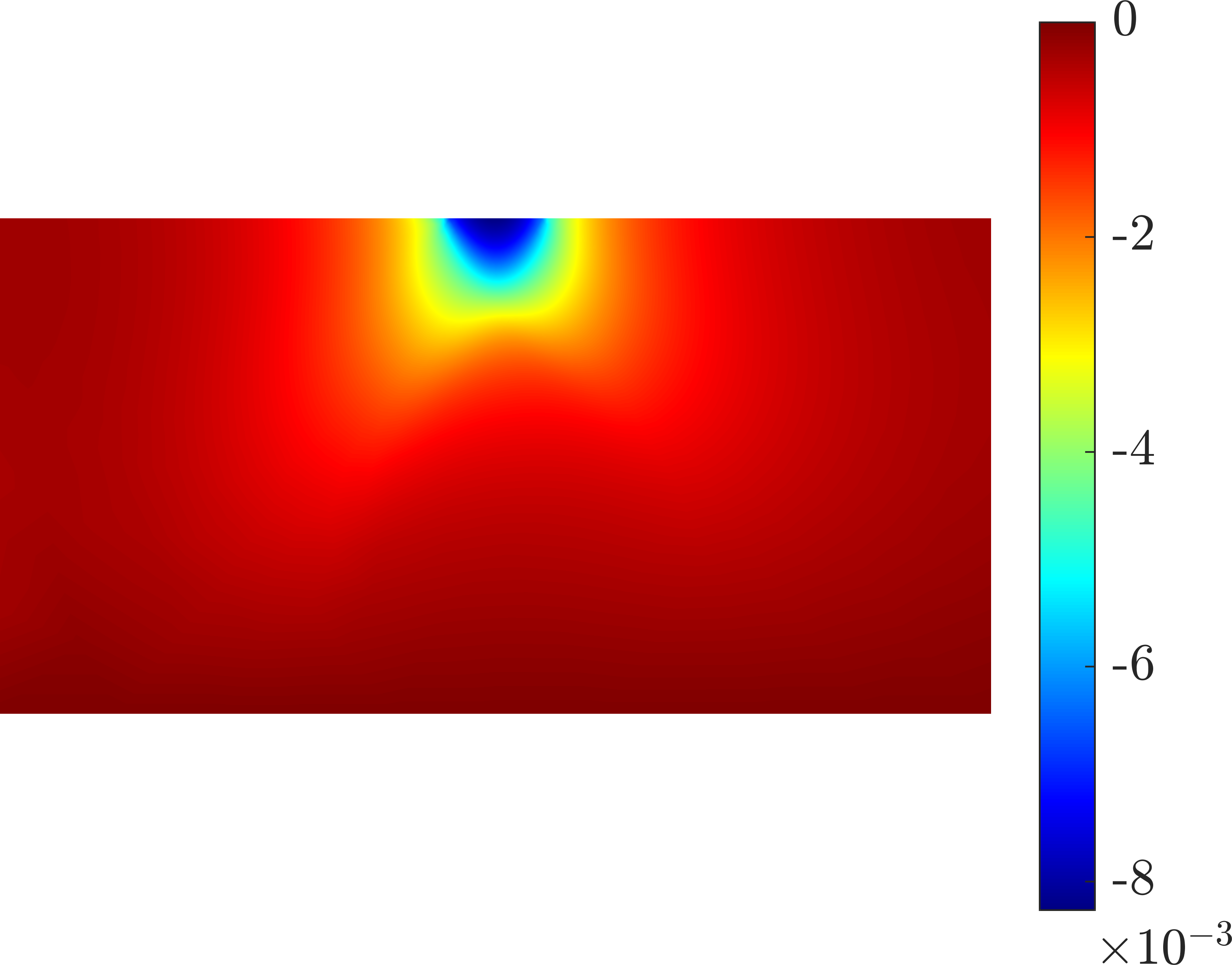} &
	\includegraphics[width = 0.3\textwidth]{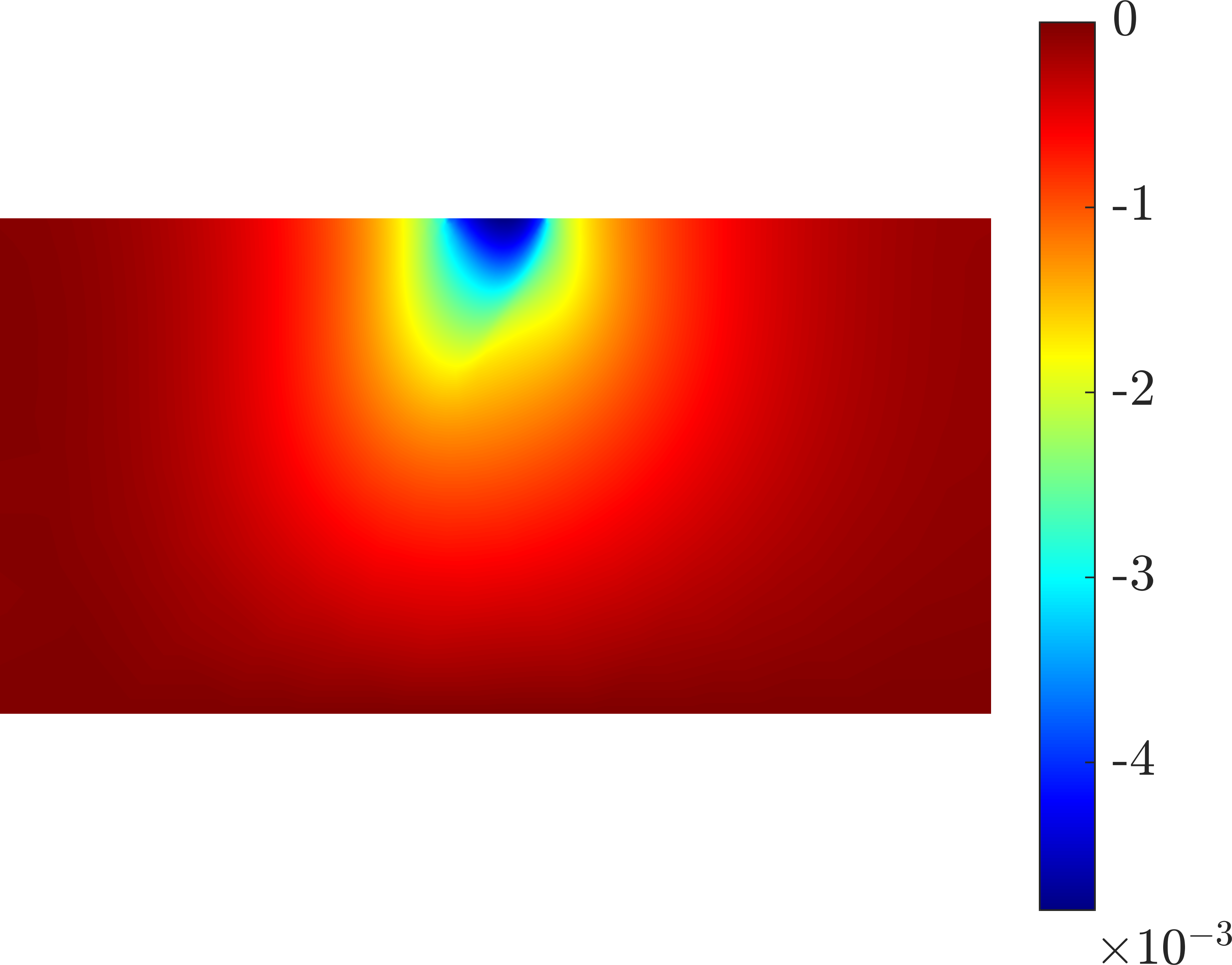} \\
	(d) $u_y$ when $\theta = \pi/15$, m & (e) $u_y$ when $\theta = \pi/4$, m & (f) $u_y$ when $\theta = 7\pi/18$, m \\
	\end{tabular}
	\end{center}
	\caption
	{\label{newcase_04} Contours of $\norm{\tens{\epsilon}^p}$ and vertical displacement $u_y$ at the end of the simulation (excess gas pressure is almost zero). Left: $\theta = \pi/15$. Middle: $\theta = \pi/4$. Right: $\theta = 7\pi/18$.}
\end{figure}

\cref{newcase_06} portrays the excess gas pressure contours at three different time slots for two different preconsolidation pressures $p_{c0} = -40$ MPa and $p_{c0} = -400$ MPa. The latter case of $p_{c0} = -400$ MPa in fact characterizes an anisotropic poroelastic material. Non-Darcy flow and $\theta = \pi/4$ are assumed for both cases. As a result, we can see that the elastic porous medium generates a lower excess gas pressure than the elastoplastic porous medium. Apart from this finding, the spatial variation of the excess gas pressure for the elastoplastic porous medium has an arch shape, as opposed to the typical bulb shape generated by the elastic porous medium. The arch shape might be attributed to the shape of the plastic zone. Both the arch shape and the bulb shape are skewed because of an inclined bedding plane orientation. The final deformation of the anisotropic elastoplastic material is greater than the anisotropic elastic material. \cref{newcase_04} depicts the final vertical displacement and norm of plastic strain tensor $\tens{\epsilon}^p$. Non-Darcy flow and $p_{c0} = -40$ MPa are assumed for all cases. We find that a change in $\theta$ would affect the shape of the skewed $\norm{\tens{\epsilon}^p}$ contours. For $\theta = \pi/4$, the contour of $\norm{\tens{\epsilon}^p}$ is skewed to the right, while for $\theta = \pi/15$, it is skewed to the left. This is because $\tens{\epsilon}^p$ is influenced by the plastic flow direction, and for the above two values of $\theta$, the corresponding plastic flow directions are opposite, as proved in Zhao et al. \cite{zhao_strength_2018}. For $\theta = 7\pi/18$, one dominant shear band that is nearly parallel to the bedding plane is generated from our simulation. The irregular patterns of $\norm{\tens{\epsilon}^p}$ for $\theta = \pi/4$ and $\theta = 7\pi/18$ are also revealed in the example of bearing capacity of anisotropic soil by using the DEM-MPM multiscale approach \cite{liang_bearing_2021}.

\begin{figure}[!htb]
	\begin{center}
	\begin{tabular}{cc}
	\includegraphics[width = 0.45\textwidth]{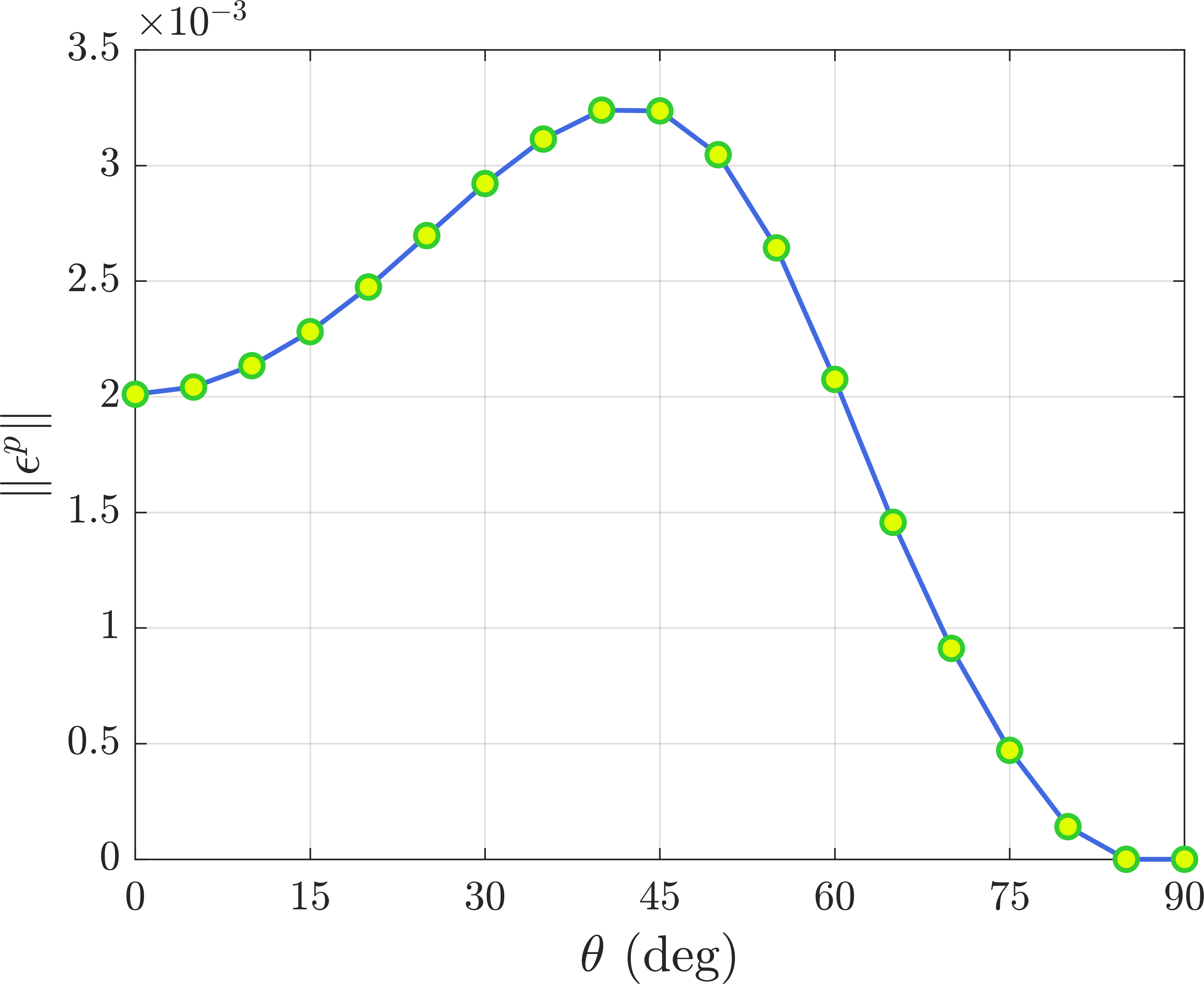} & \includegraphics[width = 0.45\textwidth]{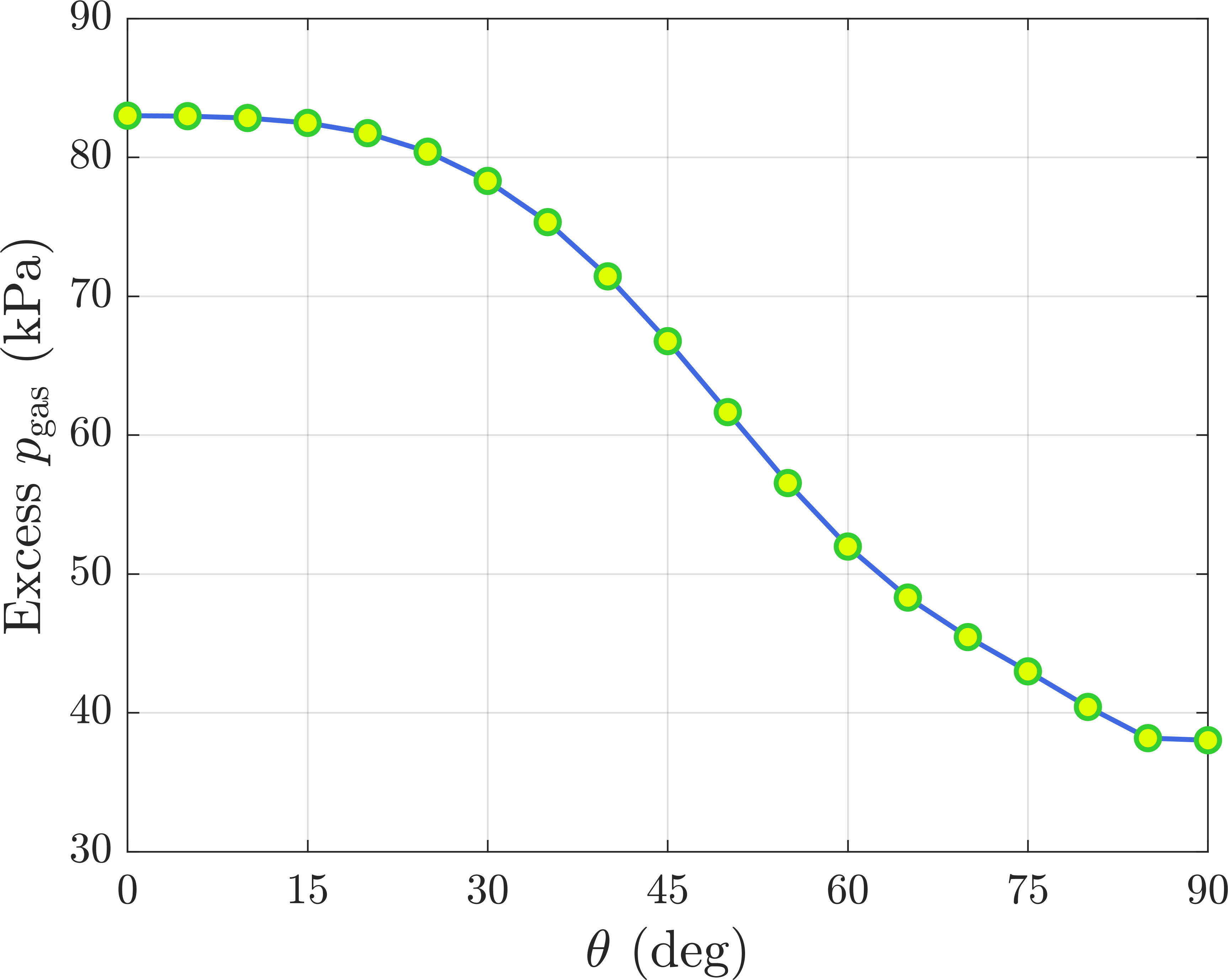}
	\end{tabular}
	\end{center}
	\caption
	{\label{newcase_08} Variation of $\norm{\tens{\epsilon}^p}$ and excess gas pressure at point A versus bedding plane orientation $\theta$ ($t = 10$ s). The left figure is consistent with the U-shaped variation of the rock strength with respect to the bedding plane orientation \citep{zhao_strength_2018}.}
\end{figure}

Finally, we observe that the plastic deformation for the medium bedding plane orientation \cref{newcase_04}(b) is greater than other two orientations, which reminds us the U-shaped variation of the rock strength with respect to the bedding plane orientation $\theta$ \cite{zhao_strength_2018}. To give a more 
intuitive result, we have plotted $\norm{\tens{\epsilon}^p}$ and $\Delta p_{\rm gas}$ (excess gas pressure) at point A with respect to $\theta$ when $t = 10$ s in \cref{newcase_08}. Now it becomes more clear that $\norm{\tens{\epsilon}^p}$ firstly increases and then decreases to zero when $\theta$ goes from 0 to $\pi/2$. The asymmetry between $\theta = 0$ and $\theta = \pi/2$ implies a higher strength in the bed-parallel direction than in the bed-normal direction. For excess gas pressure, it monotonically decreases as $\theta$ goes from 0 to $\pi/2$.

\section{Closure}

By integrating the anisotropic elasticity model and the advanced anisotropic elastoplasticity model into conventional gas production and strip footing problems, this study has provided valuable insights into the unique characteristics of unconventional shale. The analysis of gas production has uncovered new stress patterns, underscoring the significance of considering material anisotropy to enhance the fitting of field data and to exercise caution when applying isotropic results to real shale gas reservoirs. Moreover, the investigation of strip footing has revealed the substantial impact of small fluid compressibility on the undrained hydromechanical response and Mandel-Cryer effect. Furthermore, the study has identified a strong correlation between the bedding plane orientation and both the excess gas pressure and plastic strain tensor. Notably, in the anisotropic elastoplastic model, an arch-shaped pressure contour emerges when the orientation angle ($\theta$) approaches $\pi/4$. These findings deepen our understanding of gas flow and solid deformation in unconventional shale, paving the way for more precise and accurate modeling of shale behavior. Through this research, we have gained valuable insights that contribute to the advancement of knowledge in the field of unconventional shale mechanics.

\section*{Acknowledgments}

This work was supported by the Hong Kong RGC Postdoctoral Fellowship Scheme (RGC Ref. No. PDFS2223-5S04), the PolyU Start-up Fund for RAPs under the Strategic Hiring Scheme (Grant No. P0043879), the National Natural Science Foundation of China (Nos. 52004321, 52034010, and 12131014), Fundamental Research Funds for the Central Universities (Grant Nos. 20CX06025A and 21CX06031A), and Natural Science Foundation of Shandong Province, China (Grant No. ZR2020QE116).

\section*{Declaration of competing interests}
The authors declare that they have no known competing financial interests or personal relationships that could have appeared to influence the work reported in this paper.

\appendix

\section{Finite element equations of the fracture flow model}
\label{appx1}

\begin{figure}[!htb]
    \centering
    \includegraphics[scale=0.8]{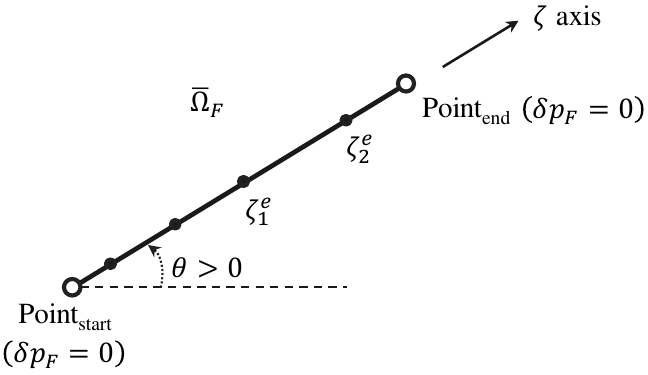}
    \caption{A schematic demonstration of the 1D fracture domain $\bar{\Omega}_F$. The domain has one starting point and one ending point, and they are regarded as the boundary of $\bar{\Omega}_F$.}
    \label{fig:A1}
\end{figure}

We consider the 1D fracture domain $\bar{\Omega}_F$, as shown in \cref{fig:A1}, which has one starting point and one ending point, and they are regarded as the boundary of $\bar{\Omega}_F$, denoted as $\partial \bar{\Omega}_F = \{\rm Point_{\,start}, \ Point_{\,end}\}$. 
In the fracture flow model, the gas pressure must be continuous at the fracture-matrix interface, and the discrete fracture elements must be located on the edges of porous matrix elements, sharing the same nodes \citep{chen_coupled_2020,chen_3d_2022}. By using this strategy, the mass exchange between the fracture and porous matrix is not required to evaluate explicitly \citep{chen_coupled_2020}, and $\partial \bar{\Omega}_F$ belongs to the type of Dirichlet boundary. Therefore, in the finite element method (FEM), the weighting function $\delta p_F$ would vanish at $\partial \bar{\Omega}_F$.

On $\bar{\Omega}_F$, a local coordinate system denoted by $\zeta$ axis is built, and the strong form of the fracture flow equation is given as
\begin{equation}
\label{FFA1}
    w \pd{\big(\phi_F\rho_g\big)}{t} + \pd{\big(\rho_g w q_F\big)}{\zeta} = 0\,,
\end{equation}
\begin{equation}
    q_F = - \frac{k_F}{\mu_g} \pd{p_F}{\zeta}\,,
\end{equation}
where $w$ is the fracture aperture, $\phi_F$ is the fracture porosity, $\rho_g$ and $\mu_g$ are calculated in the same way as before, $k_F$ is the fracture permeability. Note that we use $p_F$ to emphasize the fracture flow, but it is just the matrix gas pressure $p$ at the fracture-matrix interface including the shared nodes. In addition, experimental results suggest that $w$, $\phi_F$, and $k_F$ could be functions of $p_F$, and some empirical relations such as the exponential function have been used to determine these parameters. The relation between the local and global coordinate systems is also mentioned here. For example, fracture flow velocity $q_F$ in the global coordinate system is decomposed as a vector given as
\begin{equation}
    \mathbf{q}_F = -\frac{k_F}{\mu_g} \underbrace{\bracs{\dfrac{\partial p_F}{\partial x}\cos\theta + \dfrac{\partial p_F}{\partial y}\sin\theta}}_{\partial p_F/\partial \zeta} \begin{bmatrix}
        \cos\theta \\
        \sin\theta
    \end{bmatrix} = q_F \begin{bmatrix}
        \cos\theta \\
        \sin\theta
    \end{bmatrix}\,.
\end{equation}
It is easy to check that this vector is aligned with the 1D fracture segment, which is intuitively consistent. Similarly, the second-order derivative $\partial^2 p_F/\partial \zeta^2$ is equivalent to
\begin{equation}
    \pd[2]{p_F}{\zeta^2} = \pd[2]{p_F}{x^2} \cos^2\theta + \pd[2]{p_F}{y^2} \sin^2\theta + 2\pd[2]{p_F}{x \partial y} \sin\theta \cos\theta\,.
\end{equation}

To obtain the weak form corresponding to Eq.~\eqref{FFA1}, we multiply Eq.~\eqref{FFA1} by the weighting function $\delta p_F$ and integrate on $\bar{\Omega}_F$, the result reads
\begin{equation}
\label{(W)}
    \int_{\bar{\Omega}_F} (\delta p_F) \ w \pd{\big(\phi_F\rho_g\big)}{t} \ \d \zeta + \int_{\bar{\Omega}_F} \delta p_F \pd{\big(\rho_g w q_F\big)}{\zeta} \ \d \zeta = 0\,.
\end{equation}
It is assumed that inside the fracture, all variables remain constant in the lateral direction (the direction perpendicular to the 1D element segment), and the aperture $w$ of the fracture appearing as a factor in the 1D integral ensures the unit consistency with the 2D integral of the porous matrix \citep{karimi-fard_numerical_2003}. By using the integration by part, the second term could be simplified as
\begin{equation}
\begin{aligned}
    \int_{\bar{\Omega}_F} \delta p_F \pd{\big(\rho_g w q_F\big)}{\zeta} \ \d \zeta
    & = (\delta p_F) \ \big(\rho_g w q_F\big) \Big|^{\rm A}_{\rm B} - \int_{\bar{\Omega}_F} \big(\rho_g w q_F\big) \pd{(\delta p_F)}{\zeta} \ \d \zeta \\
    & = - \int_{\bar{\Omega}_F} \pd{(\delta p_F)}{\zeta} \big(\rho_g w q_F\big) \ \d \zeta\,,
\end{aligned}
\end{equation}
where $\rm A$ and $\rm B$ in $(\cdot)\Big|^{\rm A}_{\rm B}$ represent the starting and ending points in $\partial \bar{\Omega}_F$, and this term vanishes because the weighting function $\delta p_F$ is zero at the Dirichlet boundary $\partial \bar{\Omega}_F$. Next, by adopting the backward Euler time integration scheme, Eq.~\eqref{(W)} could be rewritten as
\begin{equation}
\label{(W)final}
    \int_{\bar{\Omega}_F} (\delta p_F) \ w \frac{\big(\phi_F\rho_g\big) - \big(\phi_F\rho_g\big)_{t_n}}{\Delta t} \ \d \zeta - \int_{\bar{\Omega}_F} \pd{(\delta p_F)}{\zeta} \big(\rho_g w q_F\big) \ \d \zeta = 0\,,    
\end{equation}
where the subscript $t_n$ implies the quantity at the previous time step, and other quantities are by default evaluated implicitly.

In this work, we assume a 1D linear element which means that for one element with local coordinates $\zeta_1^e$ and $\zeta_2^e$, the shape function $[N_F^e]$ and its derivative in the local coordinate system $[E_F^e]$ are
\begin{equation}
    [N_F^e] = \bracm{\frac{\zeta - \zeta_2^e}{\zeta_1^e - \zeta_2^e}, \quad \frac{\zeta_1^e - \zeta}{\zeta_1^e - \zeta_2^e}}\,, \quad [E_F^e] = \bracm{\frac{1}{\zeta_1^e - \zeta_2^e}, \quad \frac{-1}{\zeta_1^e - \zeta_2^e}}\,.
\end{equation}
Through the assembly operation of the element shape functions and derivatives, the global matrix equation in the residual form derived from Eq.~\eqref{(W)final} is given as
\begin{equation}
\label{resF}
    \mathcal{R}_F = \int_{\bar{\Omega}_F} [N_F]^T \ w \frac{\big(\phi_F\rho_g\big) - \big(\phi_F\rho_g\big)_{t_n}}{\Delta t} \ \d \zeta - \int_{\bar{\Omega}_F} [E_F]^T \big(\rho_g w q_F\big) \ \d \zeta\,.    
\end{equation}
By using the global $[N_F]$ and $[E_F]$ matrices, we have $p_F = [N_F][p_F]$ and $\partial p_F/\partial \zeta = [E_F][p_F]$. The size of $[N_F]$ (row vector), $[E_F]$ (row vector), and $[p_F]$ (column vector) is equal to the number of nodes (or the number of elements plus one) in $\bar{\Omega}_F$. The residual $\mathcal{R}_F$ is generally nonlinear with respect to the unknown nodal pressure vector $[p_F]$ and is best solved using Newton's method \citep{zhao_continuum_2020}. As a result, the algorithmic tangent operator $\mathcal{K}_F$ is given as
\begin{equation}
    \mathcal{K}_F = \int_{\bar{\Omega}_F} [N_F]^T\frac{\big(\phi_F w \rho_g \big)'}{\Delta t} [N_F] \ \d \zeta + \int_{\bar{\Omega}_F} [E_F]^T \chi [E_F] \ \d \zeta + \int_{\bar{\Omega}_F} [E_F]^T \pd{p_F}{\zeta} \pd{\chi}{p_F} [N_F] \ \d \zeta\,,
\end{equation}
where $()'$ implies a partial derivative with respect to $p_F$, and $\chi = \rho_g w k_F /\mu_g$. Note that the $\mathcal{R}_F$ and $\mathcal{K}_F$ should be superimposed to the finite element equations of the porous matrix \citep{chen_coupled_2020}, in order to satisfy the solvability requirements.

\bibliography{ref_short}

\end{document}